\documentclass{aa}
\usepackage{aalongtable}
\usepackage{psfig}
\usepackage{color}
\usepackage{natbib}

\def\h2{H{\small II}}

\newcounter{qub}
\begin{document}

\title{VLT/GIRAFFE spectroscopic observations of the metal-poor blue
compact dwarf galaxy SBS 0335--052E\thanks{Based on VLT observations
collected at the European Southern Observatory, Chile, Swiss GTO program 072.B-0237.}\thanks{Tables \ref{tab2} and \ref{tab4} are available only in the 
electronic version of the paper.}
}

\author{Y. I. Izotov \inst{1}
\and D. Schaerer \inst{2,3}
\and A. Blecha \inst{2}
\and F. Royer \inst{4}
\and N. G. Guseva \inst{1}
\and P. North \inst{5}}
\offprints{Y. I. Izotov, izotov@mao.kiev.ua}
\institute{          Main Astronomical Observatory,
                     Ukrainian National Academy of Sciences,
                     Zabolotnoho 27, Kyiv 03680,  Ukraine
\and
      Observatoire de Gen\`eve, 
                 51, Ch. des Maillettes,
                 1290 Sauverny, Switzerland
\and
        Laboratoire d'Astrophysique Toulouse-Tarbes, UMR 5572, 14, Av. E. Belin,
        31400 Toulouse, France
\and
 GEPI, CNRS UMR 8111, Observatoire de Paris, 5 place Janssen, 92195 Meudon Cedex, France
\and
        Laboratoire d'Astrophysique, \'Ecole Polytechnique F\'ed\'erale de Lausanne (EPFL), 
Observatoire, 1290 Sauverny, Switzerland}

\date{Received \hskip 2cm; Accepted}

\abstract
{}
{We present two-dimensional spectroscopy of the extremely 
metal-deficient blue compact dwarf (BCD) galaxy SBS 0335--052E aiming to study
physical conditions, element abundances and kinematical properties of the
ionised gas in this galaxy.}
{Observations were obtained in the spectral
range $\lambda$3620 -- 9400\AA\  with
the imaging spectrograph GIRAFFE installed on the UT2 of the Very
Large Telescope (VLT). These observations are the first ones carried out
so far with GIRAFFE in the ARGUS mode which allows to obtain  
simultaneously 308 spectra covering a 11\farcs4$\times$7\farcs3 region.}
{We produced images of SBS 0335--052E
in the continuum and in emission lines of different stages of excitation.
We find that while the maximum of emission in the majority of lines, including
the strong lines H$\beta$ 4861\AA, H$\alpha$ 6563\AA, [O {\sc iii}] 4363,5007\AA,
[O {\sc ii}] 3726,3729\AA, coincides with the youngest south-eastern star
clusters 1 and 2, the emission of He II 4686\AA\ line is offset to the
more evolved north-west clusters 4, 5. This suggests that hard ionising 
radiation responsible for the He II $\lambda$4686\AA\ 
emission is not related
to the most massive youngest stars, but rather is connected with fast radiative
shocks. This conclusion is supported by the kinematical properties of the
ionised gas from the different emission
lines as the velocity dispersion in the He II $\lambda$4686\AA\ line is
systematically higher, by $\sim$ 50\% -- 100\%, than that in other lines.
The variations of the emission line profiles suggest the presence of an
ionised gas outflow in the direction perpendicular to the galaxy disk.
We find a relatively high electron number density $N_e$ of several hundred 
cm$^{-3}$ in the brightest part of SBS 0335--052E.
There is a small gradient of the electron temperature $T_e$ and oxygen
abundance from the East to the West with systematically higher $T_e$ and lower
12+log O/H in the western part of the galaxy. The oxygen abundances for the
whole H II region and its brightest part are 12 + log O/H = 
7.29 $\pm$ 0.02 and 7.31 $\pm$ 0.01, respectively. We derive the He mass 
fraction taking into
account all systematic effects. The He mass fraction $Y$ = 
0.2463 $\pm$ 0.0030, derived from the emission of the whole H II region,
is consistent with the primordial value predicted by the standard big bang 
nucleosynthesis model. We confirm the presence of Wolf-Rayet stars in the cluster 3.}
{}

\authorrunning{Izotov et al.}

\titlerunning{VLT/GIRAFFE spectroscopic observations of SBS 0335--052E}

\maketitle

\section{Introduction}

\begin{figure}
\hspace*{1.0cm}\psfig{figure=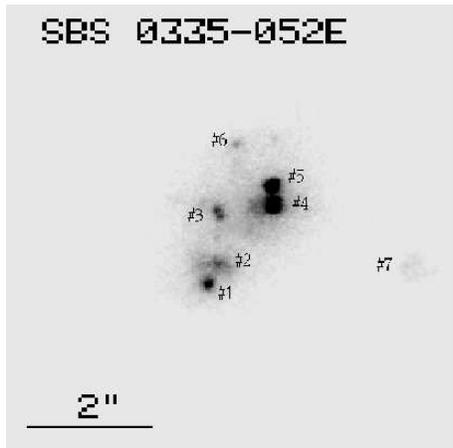,angle=0,width=6.0cm}
\caption{Archival HST/ACS UV image of SBS 0335--052E with the labelled
compact clusters.
North is up and East is to the left.
}
\label{fig1}
\end{figure}

The blue compact dwarf (BCD) galaxy SBS 0335--052E is an excellent nearby 
laboratory for studying star formation in low-metallicity environments.
Since its discovery as one of the
most metal-deficient star-forming galaxies known \citep{I90}, 
with oxygen abundance 12 + log O/H $\sim$ 7.30 \citep{M92,I97b,I99,TI05}, 
SBS 0335--052E has often been proposed as a nearby young dwarf 
galaxy \citep{I90,I97b,T97,P98,P04}. \citet{T97} and \citet{P98}, using the 
same Hubble Space Telescope (HST) images, have found several luminous clusters.
The brightest clusters are labelled in Fig. \ref{fig1} 
which represents the highest spatial resolution archival UV HST/Advanced 
Camera for Surveys (ACS) image of SBS 0335--052E obtained by \citet{K03}.
Some of the clusters are very young and produce extended regions of ionised gas
\citep{M92,I97b,P98,P04}. In particular, \citet{I01b}, using deep long-slit 
spectra of SBS 0335--052E,
have shown that extended H$\alpha$ emission is detected over
$\sim$ 6--8 kpc, suggesting that hot ionised gas is spread out far away from 
the central part of the galaxy.

\begin{table*}[t]
\caption{Journal of observations \label{tab1}}
\begin{tabular}{lccrccc} \hline
Date       & Setup & Wavelength& Resolving &Exposure,     & Airmass$^{\rm a}$& Seeing$^{\rm a}$   \\ 
           &       & range, \AA& power $R$ &   s        &        &    \\ 
\hline
19 Nov 2003& LR1 & 3620--4081       &12800&2$\times$1200 &1.066, 1.065    &1\farcs3, 0\farcs9  \\
20 Nov 2003& LR2 & 3964--4567       &10200&2$\times$1200 &1.082, 1.062    &0.8, 0.8      \\
20 Nov 2003& LR3 & 4501--5078       &12000&2$\times$1350 &1.061, 1.080    &1.0, 0.7      \\
21 Nov 2003& LR4 & 5010--5831       & 9600&2$\times$1200 &1.295, 1.170    &0.8, 0.8      \\
21 Nov 2003& LR5 & 5741--6524       &11800&2$\times$1500 &1.070, 1.066    &1.3, 0.7      \\
21 Nov 2003& LR6 & 6438--7184       &13700&2$\times$1200 &1.067, 1.103    &1.0, 0.7      \\
21 Nov 2003& LR7 & 7102--8343       & 8900&2$\times$1200 &1.286, 1.486    &0.8, 0.9      \\
22 Nov 2003& LR8 & 8206--9400       &10400&2$\times$1200 &1.076, 1.061    &0.7, 0.8      \\ \hline
\end{tabular}

$^{\rm a}$The first value is at start of exposure, the second value is at end of exposure.

\end{table*}

\citet{TI97} using HST/GHRS UV spectrum of SBS 0335--052E have discovered
a very broad Ly$\alpha$ line in absorption suggesting that this galaxy is 
embedded in a large envelope of neutral gas. The column density of
$N$(H {\sc i}) = 7$\times$10$^{21}$ cm$^{-2}$ in SBS 0335--052E 
derived by \citet{TI97} is the largest one known for the BCDs. Later,
\citet{P01} using Very Large Array (VLA) observations in the line H {\sc i}
$\lambda$21 cm have detected a large neutral gas cloud around SBS 0335--052E 
with a size 66 by 22 kpc elongated in the east-west direction
and with two maxima separated by 22 kpc. The first
maximum in H {\sc i} distribution is connected to SBS 0335--052E, and the
second one to the companion dwarf galaxy SBS 0335--052W discovered 
by \citet{P97}. The latter galaxy is shown by \citet{I05a} to be
the lowest metallicity emission-line galaxy known with 
12 + log O/H = 7.12 $\pm$ 0.03. \citet{T05} using Far Ultraviolet Spectroscopic
Explorer (FUSE) observations have found that the oxygen abundance of the 
neutral gas around SBS 0335--052E of 12 + log O/H $\sim$ 7.0 is only slightly
lower than that of the ionised gas, implying that this galaxy was not formed
from pristine gas. 

Despite the low metallicity of SBS 0335--052E, an appreciable amount of dust
has been detected in it. \citet{I97b} and \citet{T97} have found variations
of extinction in this galaxy from the optical spectroscopic and photometric
observations. Later, \citet{T99} and \citet{H04} using 
Infrared Space Observatory (ISO) and Spitzer 
mid-infrared observations have found an emission of an appreciable amount of
warm dust with a characteristic temperature of $\sim$ 100K. Even hotter
dust with a temperature of several hundred degrees is expected to be present 
in SBS 0335--052E which is indicated in the infrared spectra at shorter 
wavelengths, namely 2--4 $\mu$m \citep{V00,H01}.

Since the discovery paper by \citet{I90} it is known that the ionising
radiation in SBS 0335--052E is hard. The presence of the high-ionisation
He {\sc ii} $\lambda$4686\AA\ emission line \citep[e.g. ][]{I90,M92,I97b}
and of the [Fe {\sc v}] $\lambda$4227\AA\ emission line \citep{I01b,F01} suggests
that radiation with photon energies greater than 4 Rydberg is intense and 
could not be explained by stellar emission. Furthermore, \citet{I01b}
and \citet{TI05}
have found [Fe {\sc vii}] and [Ne {\sc v}] emission lines which require 
intense radiation with photon energies above 7 Rydberg. SBS 0335--052E has
also been detected in the X-ray range \citep{T04}. The origin of the hard
ionising radiation remains unclear. Several mechanisms such as radiation
from most massive main-sequence stars, Wolf-Rayet stars, high-mass X-ray 
binaries and radiative shocks have been discussed
by e.g. \citet{G91}, \citet{S96}, \citet{I01b}, \citet{I04} and \citet{TI05}. 
The most recent investigations have shown
that although the stellar origin
of hard radiation is not completely excluded, the most likely source of hard
radiation is fast radiative shocks.

Despite the efforts of different groups in studying the properties of 
SBS 0335--052E and its evolutionary status many problems remain unsolved.
Since this galaxy is possibly a young galaxy it could be considered as a local
counterpart of the high-redshift young dwarf galaxies.
Therefore the continuation of
its studies is important for cosmological applications. 
In this paper we present a two-dimensional
spectroscopic study of SBS 0335--052E with the VLT/GIRAFFE. These are
the first observations carried out so far in the ARGUS mode. 
Two new features
are characteristic for these new observations which were not present in all
previous spectroscopic studies of this galaxy. First, new observations allow
to map the whole galaxy in different emission lines and in the continuum. This gives
integrated characteristics of different emission lines for the whole H {\sc ii}
region, such as the line luminosities, which are necessary input parameters
for building up the model of the H {\sc ii} region. Second,
the spectral resolution of new observations is by one order of magnitude
better than in all previous spectroscopic observations of SBS 0335--052E
and it is enough to obtain the intrinsic profiles of emission lines.
This allows to study the kinematics of the H {\sc ii} region, and to make a
comparison of kinematic characteristics in regions of different ionisation
stages. In particular, it is important to compare the 
He {\sc ii} $\lambda$4686\AA\ 
line profiles with the profiles of other emission lines and, hence, to make
conclusions concerning the origin of hard radiation in SBS 0335--052E.

In \S\ref{sec:obs} we describe the observations and data reduction. Morphology
of SBS 0335--052E in different emission lines and continuum is considered in
\S\ref{sec:morph}. Kinematic properties are discussed in \S\ref{sec:vel}.
Heavy element abundances and helium abundance are derived in \S\ref{sec:abund}.
The Wolf-Rayet stellar population in SBS 0335--052E is discussed in
\S\ref{sec:wr}.
Our conclusions are summarised in \S\ref{sec:concl}.

\section{Observations and Data Reduction \label{sec:obs}}

Observations of SBS 0335--052E with the VLT/GIRAFFE spectrograph have been 
done during the nights 19 -- 22 November, 2003 in the entire visible range. 
GIRAFFE is equipped with a 2K$\times$4K EEV CCD. The size of the CCD pixels is
15 $\mu$m$\times$15 $\mu$m. The spatial scale is 
0\farcs52/pixel in ARGUS direct mode which was used during our observations.
The ARGUS array is a rectangular array of 22 by 14 microlenses which is fixed 
at the center of one positioner arm. We used a spatial scale with a sampling 
of 0\farcs52 per microlens and a total aperture of 11\farcs4$\times$7\farcs3. 
The major axis of the array was directed along the major axis of SBS 0335--052E
at a position angle P.A. = --30$^\circ$ and centered on the cluster 1
(Fig. \ref{fig1}). The low-resolution mode with the 600 lines/mm grating
has been used. Since the wavelength coverage with this grating
ranges from 400\AA\ to 1200\AA\ depending on the central wavelength, eight exposures
have been done with 
setups LR1 -- LR8 (Table \ref{tab1}) to obtain panoramic spectra in the
wavelength range $\lambda$3620 -- 9400\AA\ and a spectral resolution of
0.5 -- 1\AA. Each exposure was split in two subexposures for removal
of cosmic ray hits. The journal of observations is shown in Table \ref{tab1}.
Additionally, for the same setups the spectra of two standard stars, Feige 110
and HD 49798, have been obtained for flux calibration. During the days the
exposures of bias, Nasmyth screen flats and comparison lamps for the wavelength 
calibration have been obtained. The description of GIRAFFE may be found
in \citet{P02}.

The spectra were extracted and calibrated using  the standard {\em Python} 
version of BLDRS~-~Baseline   Data    Reduction   Software   (girbldrs-1.12) 
available from {\em http://girbldrs.sourceforge.net}. Basic description of 
BLDRS is given in \citet{Bla_00} and \citet{R02}. 
The processing includes the bias subtraction,
correction for the pixel sensitivity variations, localisation, optimal 
extraction, rebinning to linear wavelength scale and night sky subtraction.

\begin{figure*}
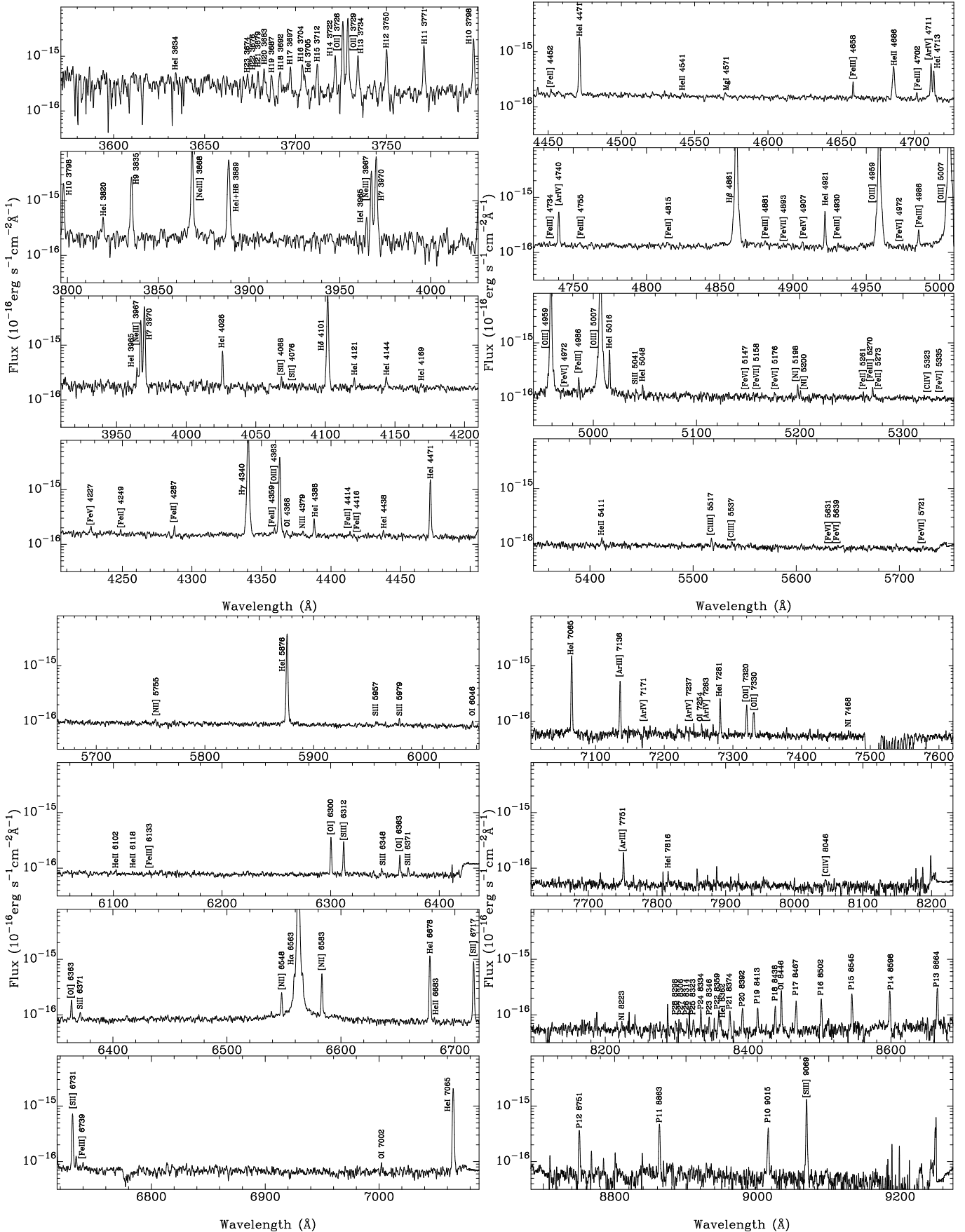

\hspace*{-0.0cm}\psfig{figure=wfig1_1l.ps,angle=0,width=9.0cm}
\hspace*{-0.0cm}\psfig{figure=wfig1_2l.ps,angle=0,width=9.0cm,clip=}
\hspace*{-0.0cm}\psfig{figure=wfig1_3l.ps,angle=0,width=9.0cm}
\hspace*{-0.0cm}\psfig{figure=wfig1_4l.ps,angle=0,width=9.0cm,clip=}
\caption{Spectrum of the brightest part of SBS 0335--052E shown in 
Fig. \ref{fig7}a as a rectangular region delineated by thick solid line. 
}
\label{fig2}
\end{figure*}

The flux calibration has been done using  IRAF\footnote{IRAF is 
the Image Reduction and Analysis Facility distributed by the 
National Optical Astronomy Observatory, which is operated by the 
Association of Universities for Research in Astronomy (AURA) under 
cooperative agreement with the National Science Foundation (NSF).}.
The flux-calibrated and redshift-corrected spectrum of the 
brightest rectangular region
delineated by the thick solid line in the center of Fig. \ref{fig7}a is shown 
in Fig. \ref{fig2}. The region includes clusters 1 and 2. Many
strong permitted and forbidden emission lines are seen in 
the spectrum of this region. One
important feature of this spectrum is that its spectral resolution is the 
highest among all other spectra of SBS 0335--052E obtained so far. In
particular, several blends are resolved for the first time for this object,
most notably, [O {\sc ii}] $\lambda$3726 and [O {\sc ii}] $\lambda$3729,
He {\sc i} $\lambda$3965, [Ne {\sc iii}] $\lambda$3967 and H7 $\lambda$3970,
[Ar {\sc iv}] $\lambda$4711 and He {\sc i} $\lambda$4713. The spectrum 
contains emission lines of ions of a broad range of ionisation stages. In
particular, [Fe {\sc ii}], [Fe {\sc iii}], [Fe {\sc iv}],
[Fe {\sc v}], [Fe {\sc vi}] and [Fe {\sc vii}] emission lines are detected.

\begin{figure*}
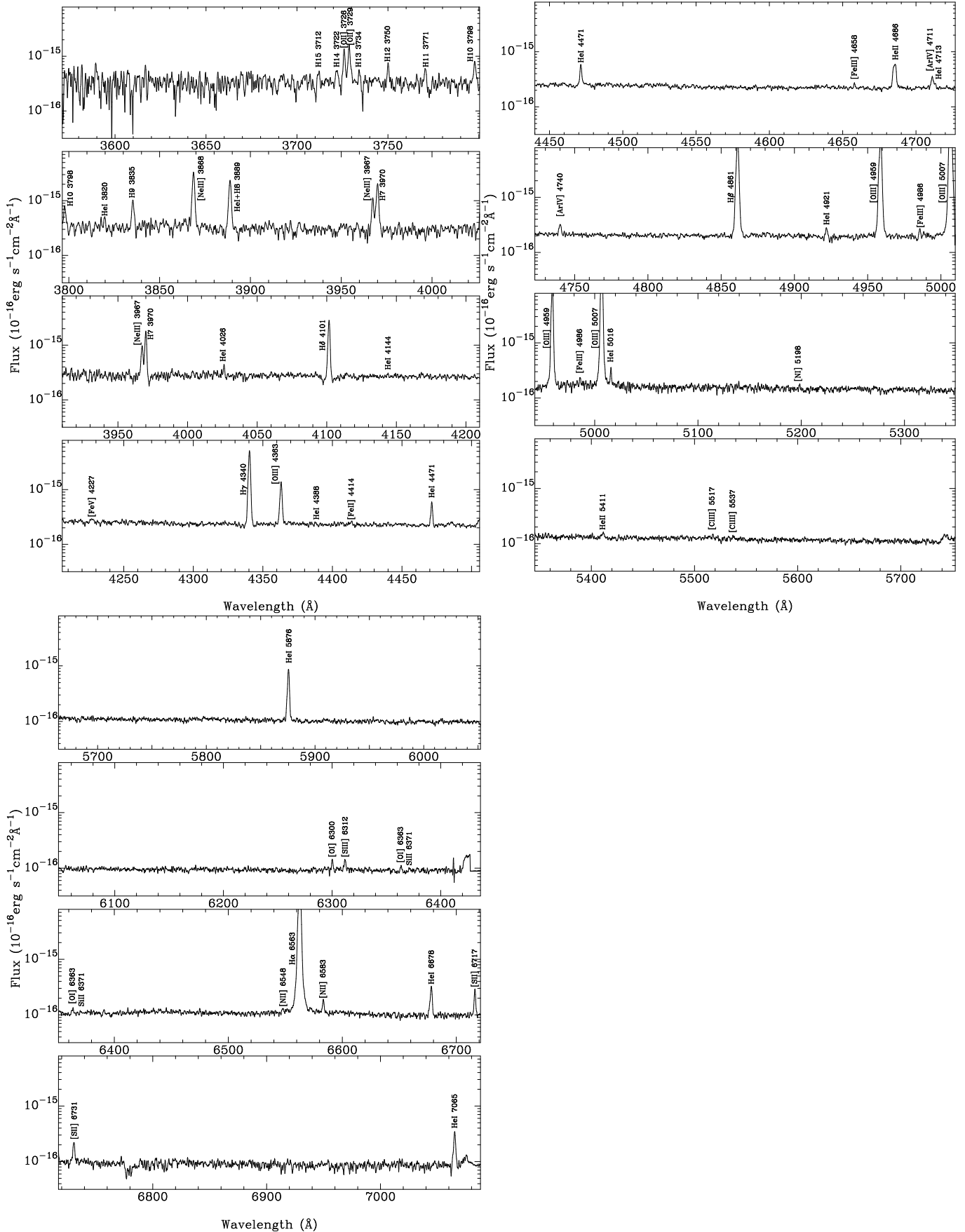

\hspace*{-0.0cm}\psfig{figure=wfig1_1w.ps,angle=0,width=9.0cm}
\hspace*{-0.0cm}\psfig{figure=wfig1_2w.ps,angle=0,width=9.0cm,clip=}
\hspace*{-0.0cm}\psfig{figure=wfig1_3w.ps,angle=0,width=9.0cm}
\caption{Spectrum of a region centered on
clusters 4 and 5 and shown in Fig. \ref{fig7}a as a square region
delineated by thick dashed line.
}
\label{fig3}
\end{figure*}

The spectrum of another region 
centered on the clusters 4 and 5 and delineated by the thick dashed
line in Fig. \ref{fig7}a is shown in
Fig. \ref{fig3}. The level of continuum in this spectrum is comparable
to that in the spectrum of the brightest region (Fig. \ref{fig2}), but 
emission lines are weaker.

The emission line fluxes and widths were measured in each of the
22$\times$14 lens array using the routine SPLOT in
IRAF. Flux errors were derived from the photon statistics using 
non-flux-calibrated spectra. These errors were propagated in the determination
of the electron temperatures, electron number densities and element
abundances.

\begin{figure*}
\hspace*{0.0cm}\psfig{figure=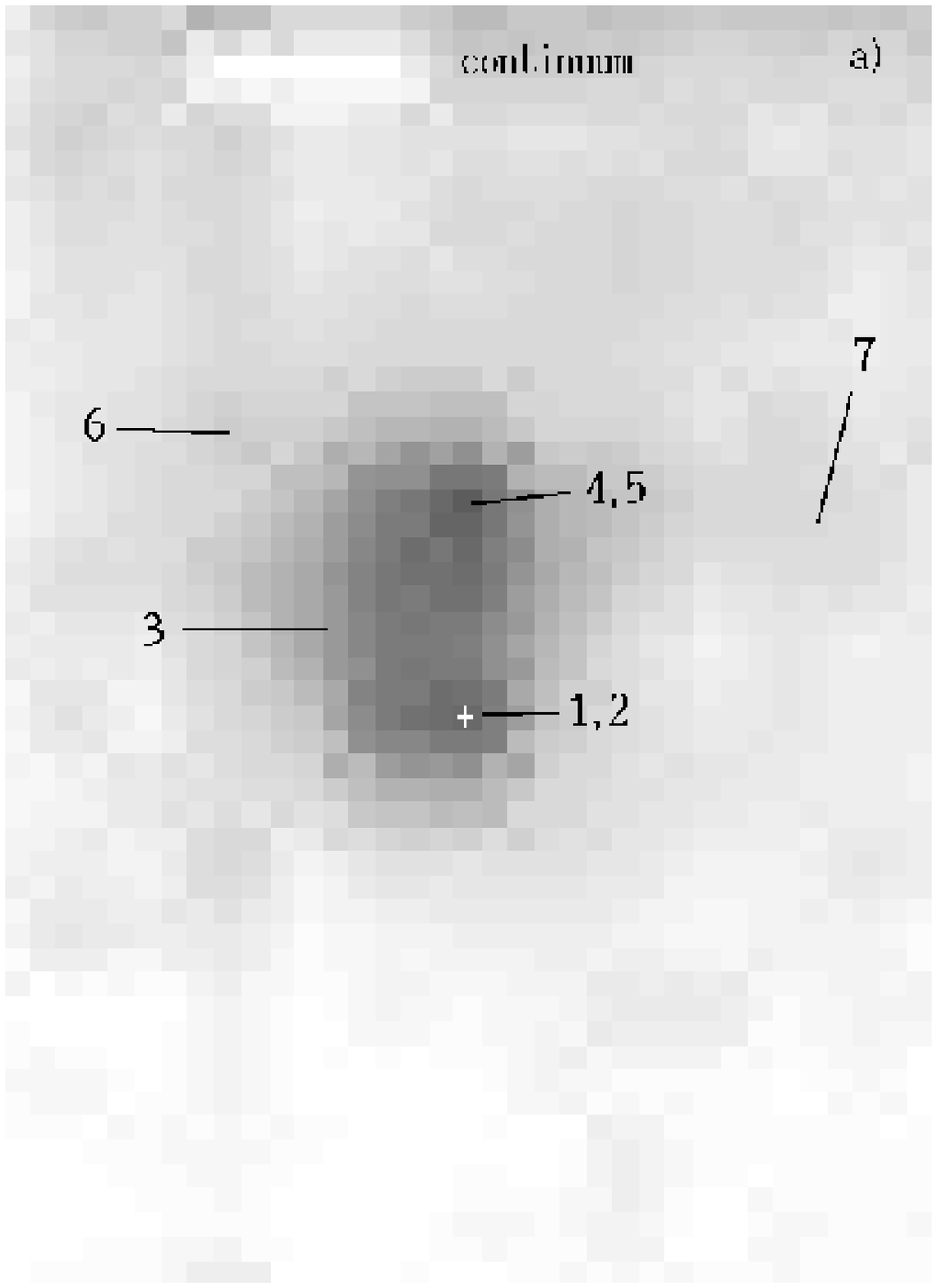,angle=0,width=3.5cm}
\hspace*{0.1cm}\psfig{figure=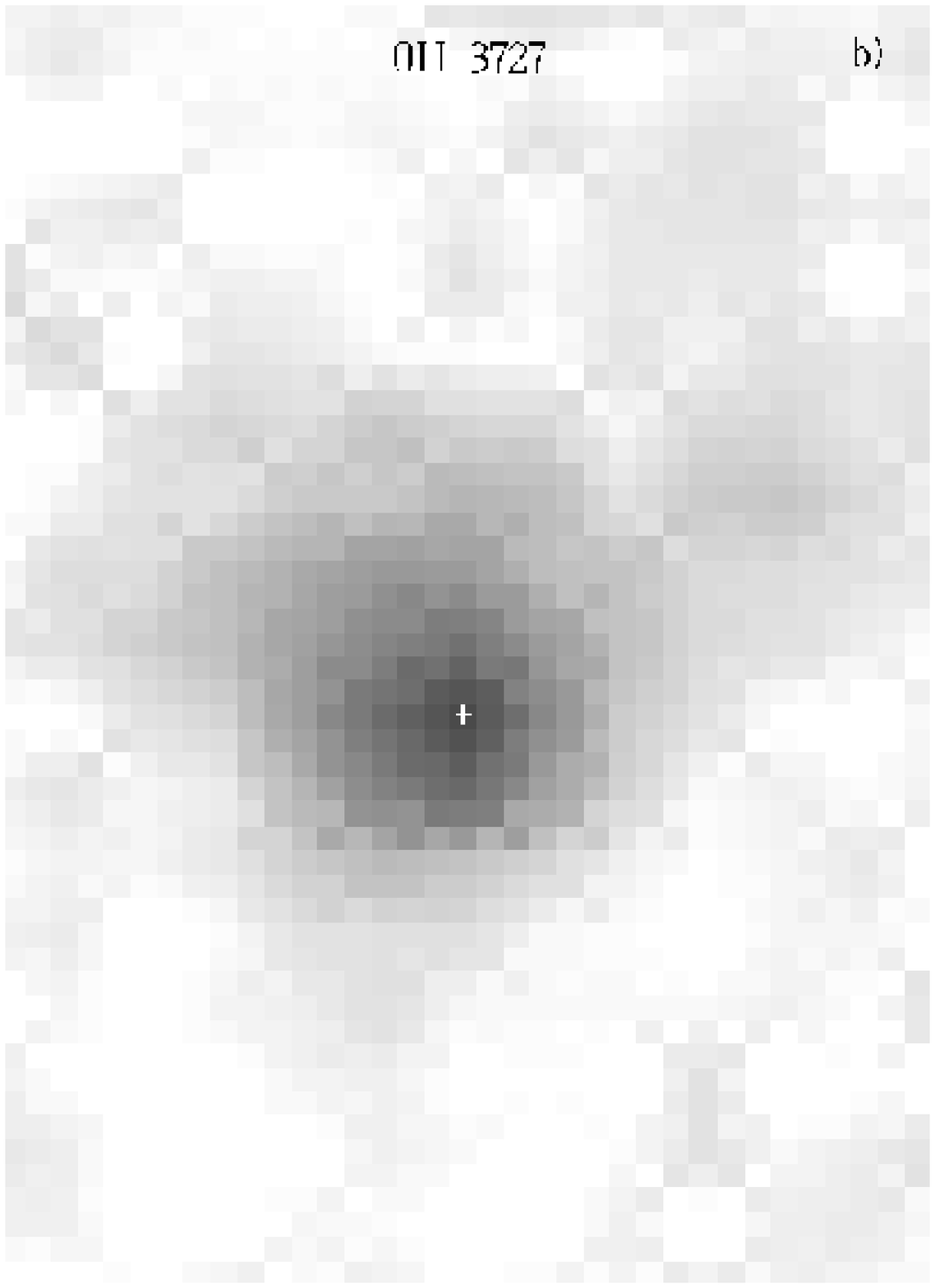,angle=0,width=3.5cm}
\hspace*{0.1cm}\psfig{figure=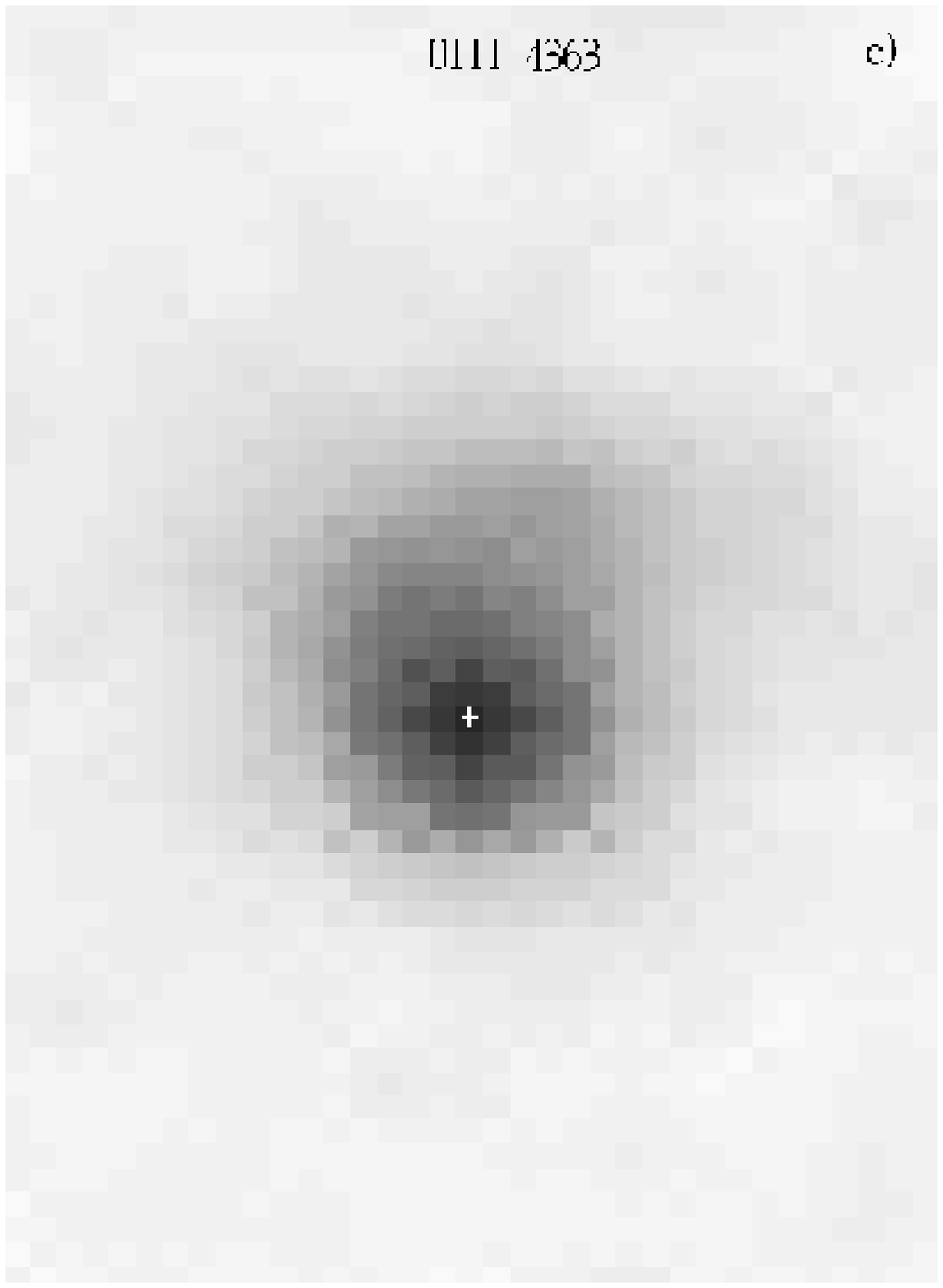,angle=0,width=3.5cm}
\hspace*{0.1cm}\psfig{figure=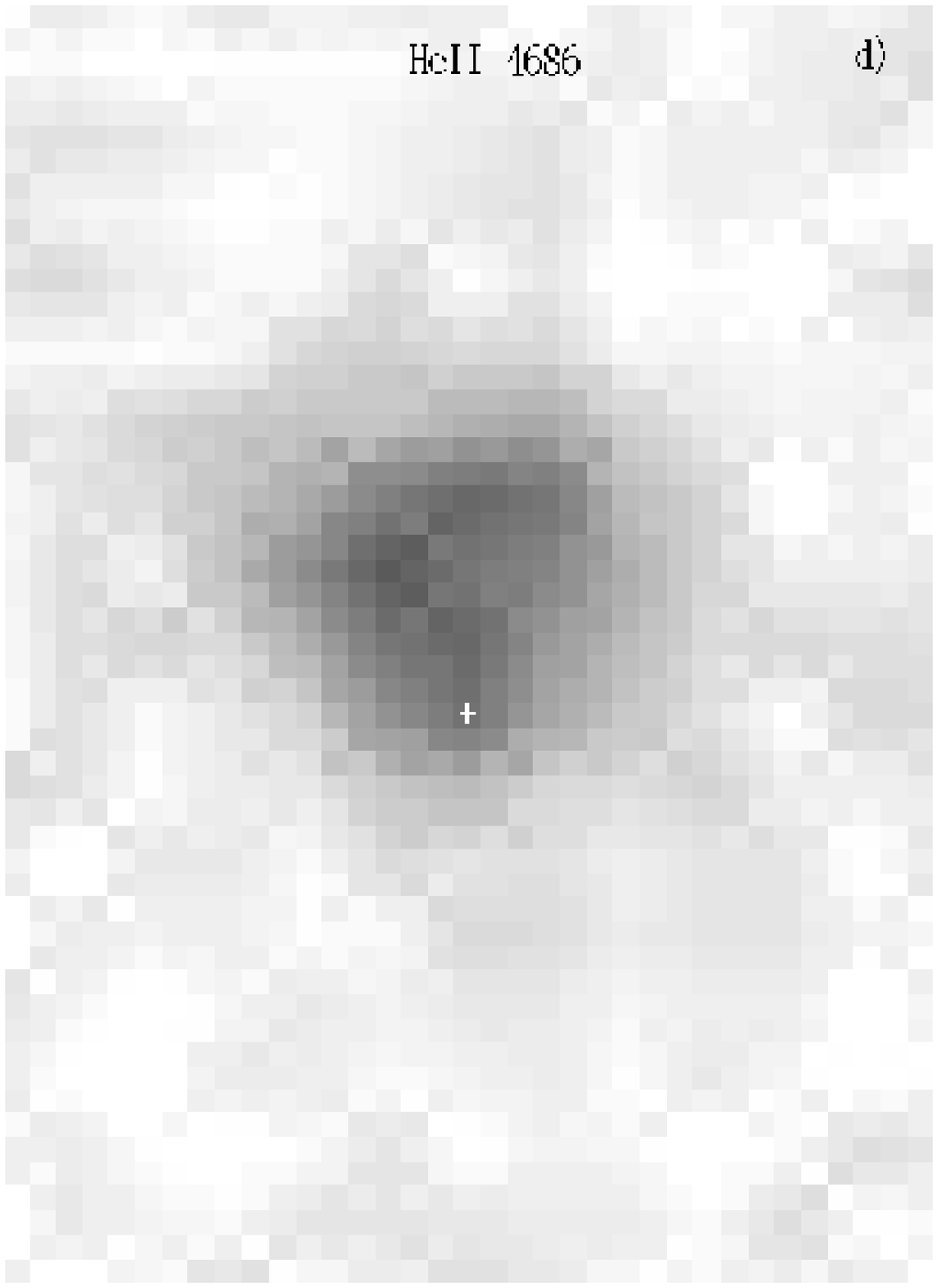,angle=0,width=3.5cm}
\hspace*{0.1cm}\psfig{figure=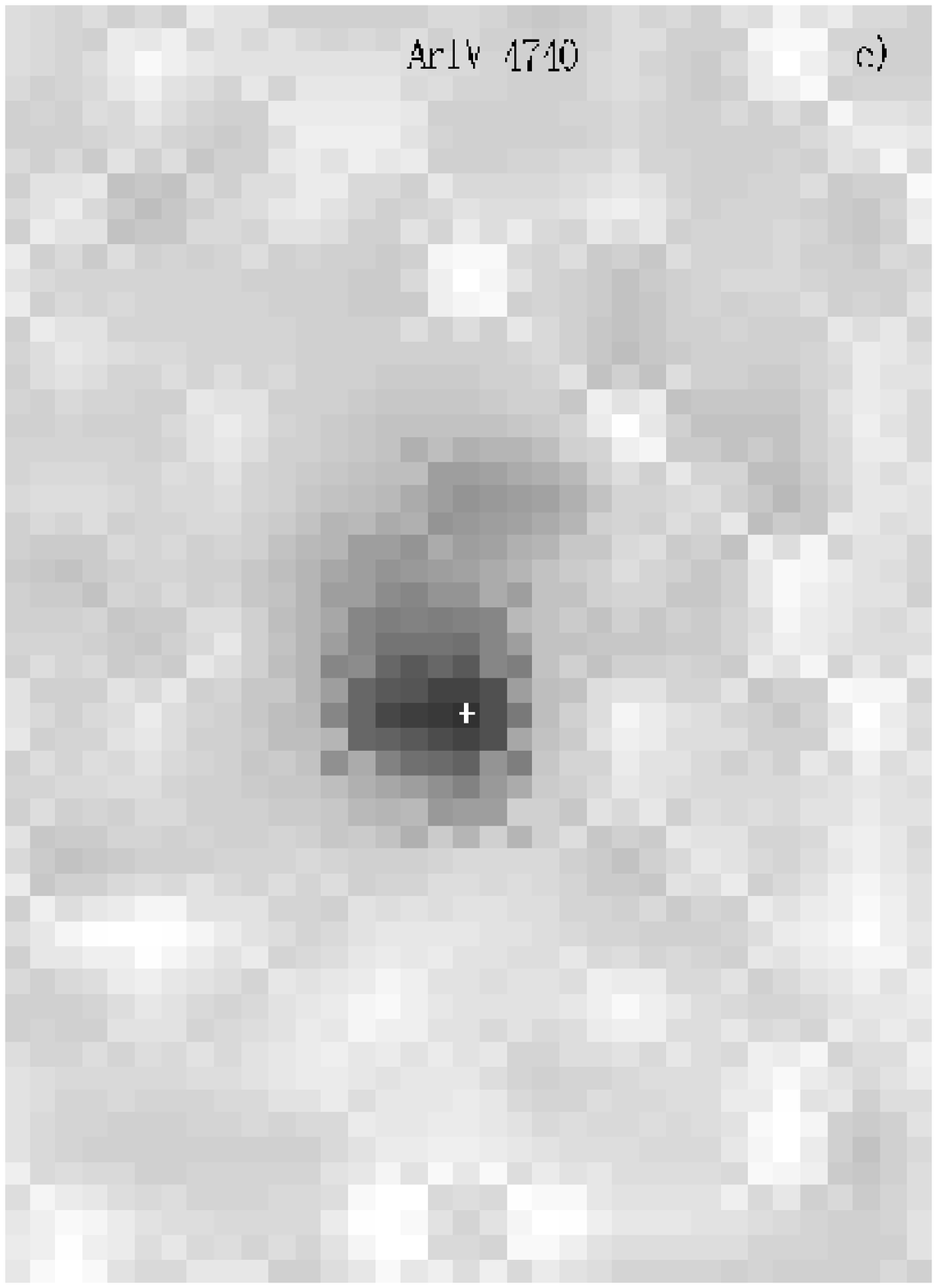,angle=0,width=3.5cm,clip=}
\hspace*{0.0cm}\psfig{figure=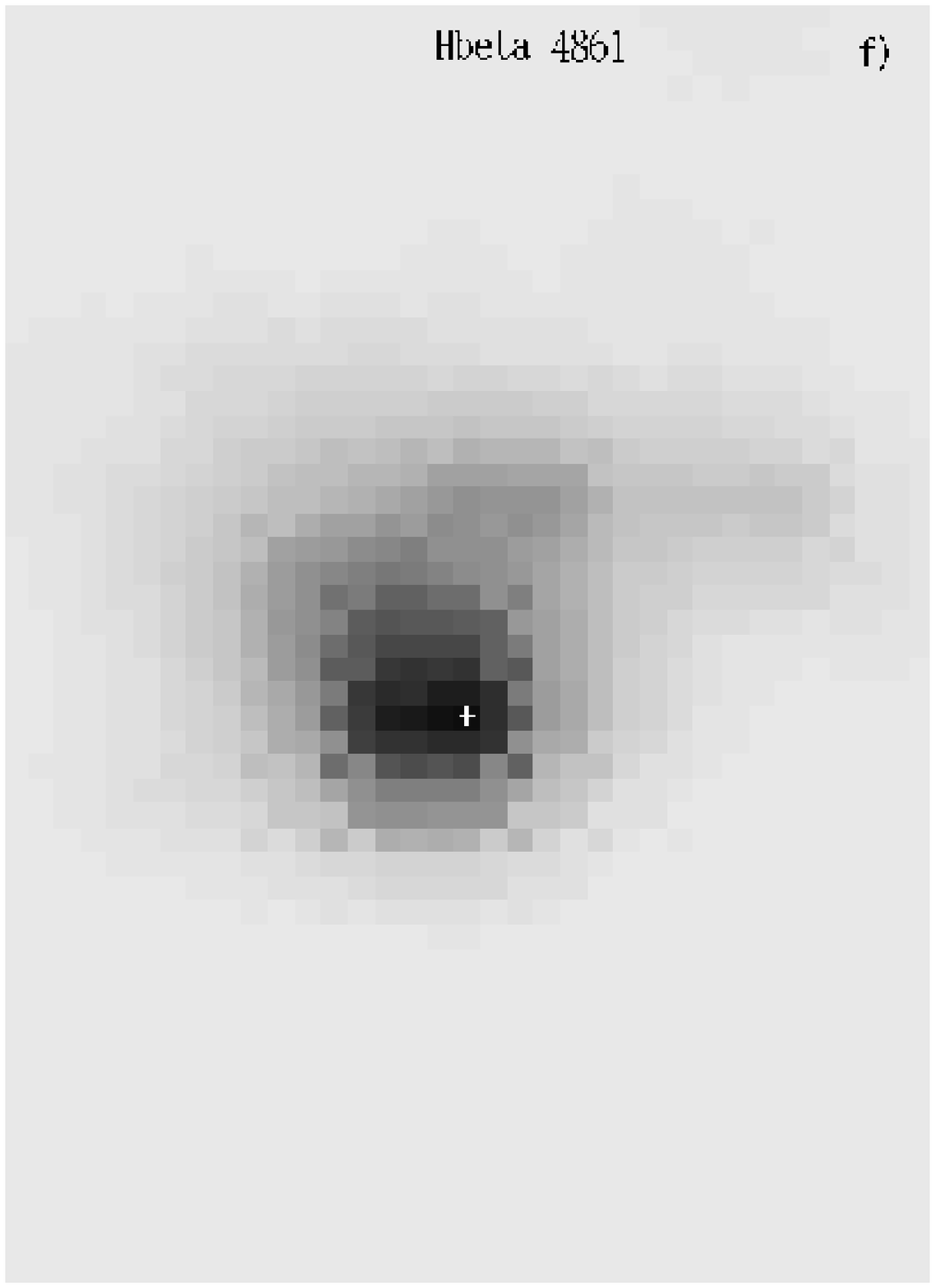,angle=0,width=3.5cm}
\hspace*{0.1cm}\psfig{figure=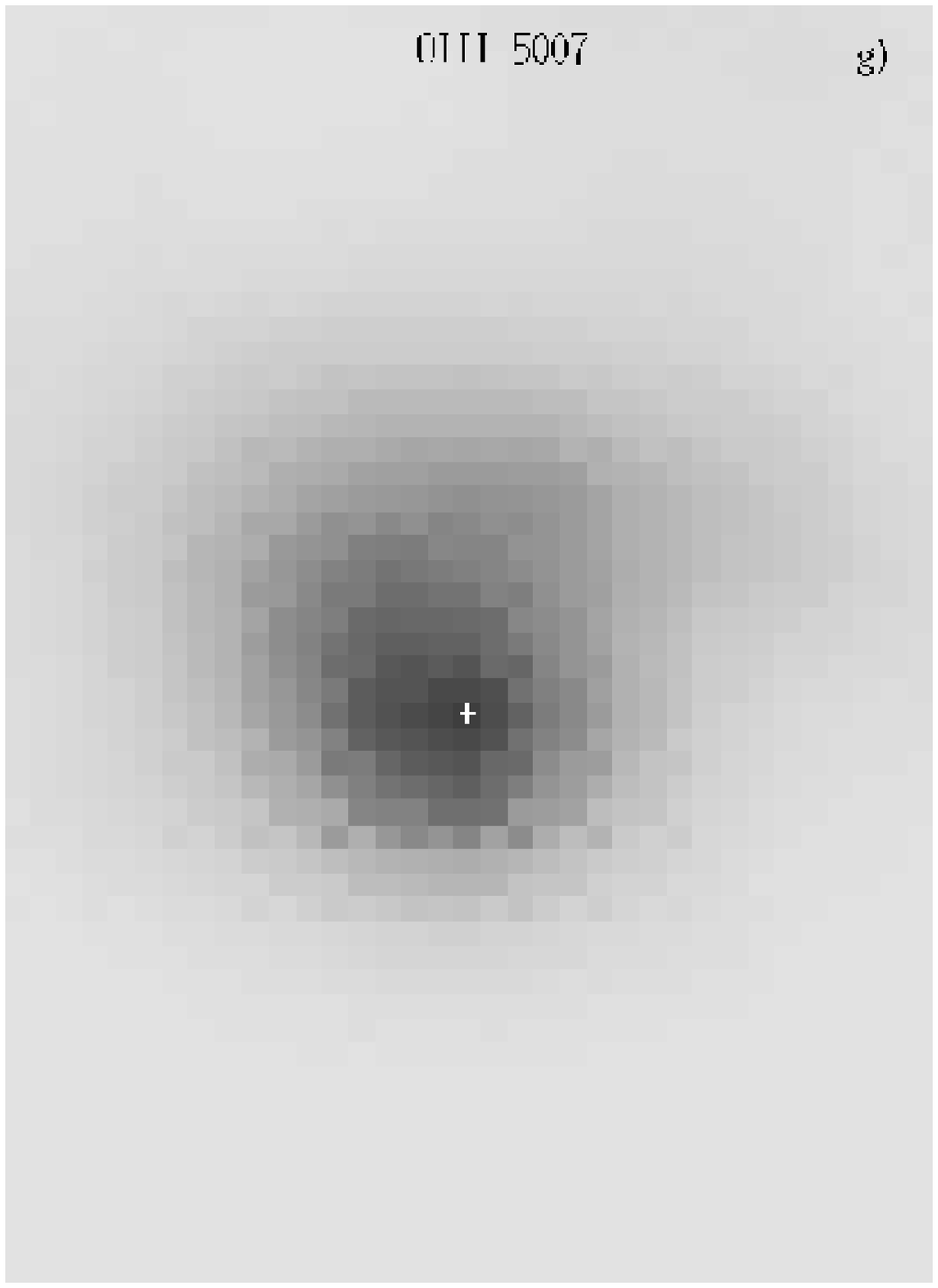,angle=0,width=3.5cm}
\hspace*{0.1cm}\psfig{figure=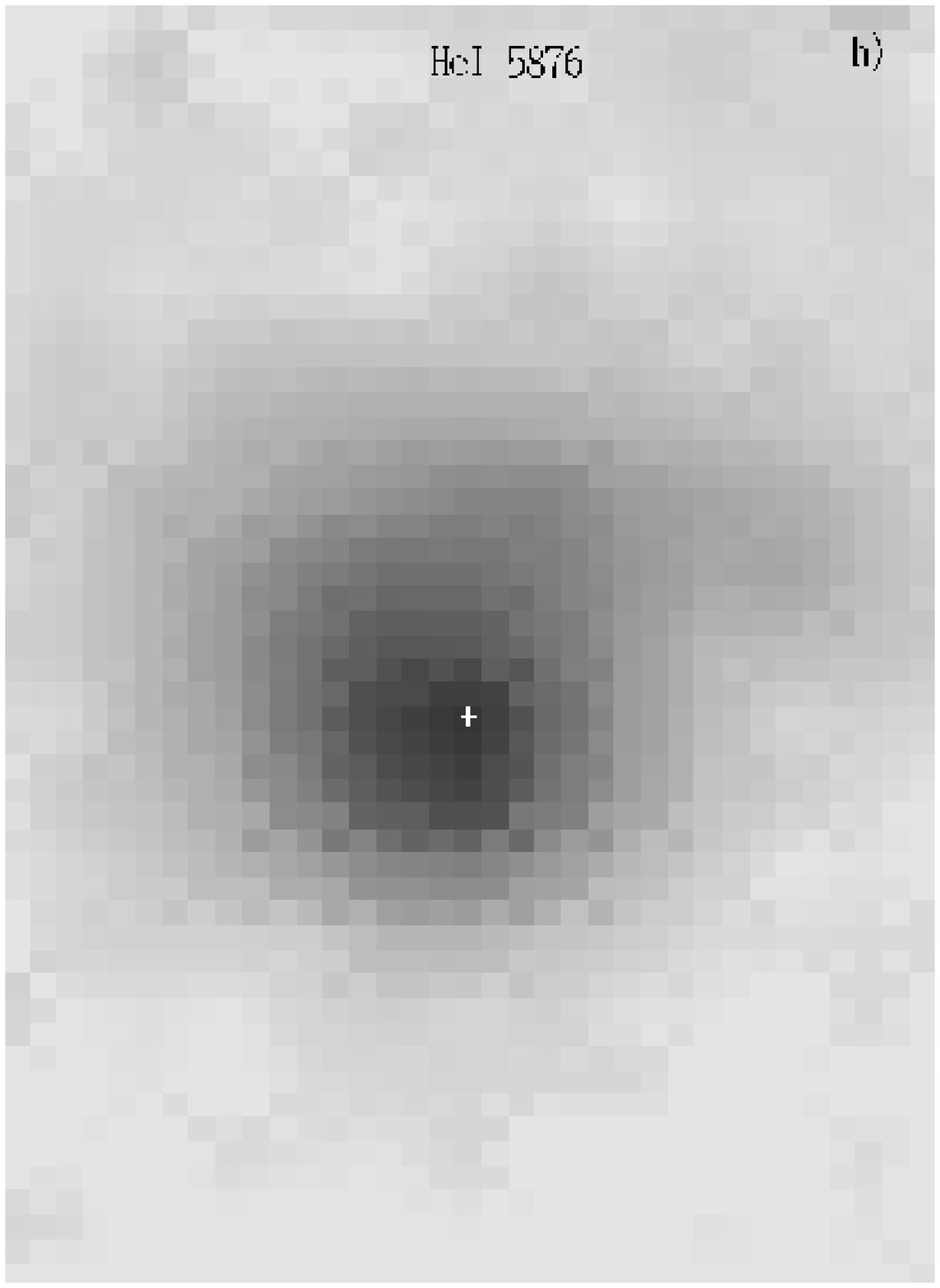,angle=0,width=3.5cm}
\hspace*{0.1cm}\psfig{figure=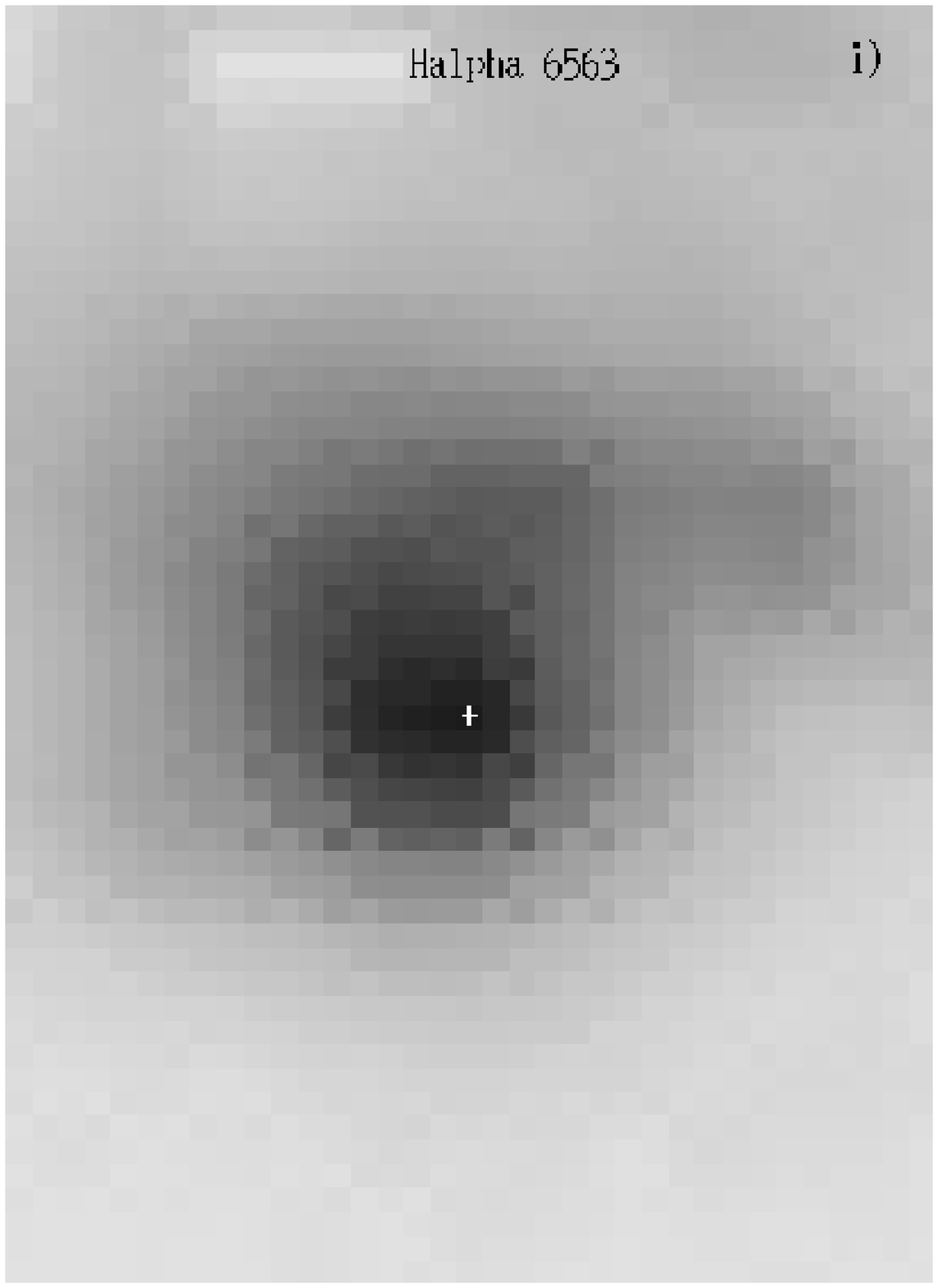,angle=0,width=3.5cm,clip=}
\caption{
Images in the continuum near H$\beta$ emission line (a) 
and in emission lines [O {\sc ii}] $\lambda$3727\AA\ (b), 
[O {\sc iii}] $\lambda$4363\AA\ (c), He {\sc ii} $\lambda$4686\AA\ (d), 
[Ar {\sc iv}] $\lambda$4740\AA\ (e),
H$\beta$ $\lambda$4861\AA\ (f), [O {\sc iii}] $\lambda$5007\AA\ (g), 
He {\sc i} $\lambda$5876\AA\ (h) and 
H$\alpha$ $\lambda$6563\AA\ (i). All images are shown in the logarithmic 
flux scale. Positions of stellar clusters are labelled in panel (a). 
White cross in each panel shows the location 
of the region with maximum flux of the H$\alpha$ $\lambda$6563\AA\ 
emission line. 
}
\label{fig4}
\end{figure*}

\begin{figure*}
\hspace*{2.0cm}\psfig{figure=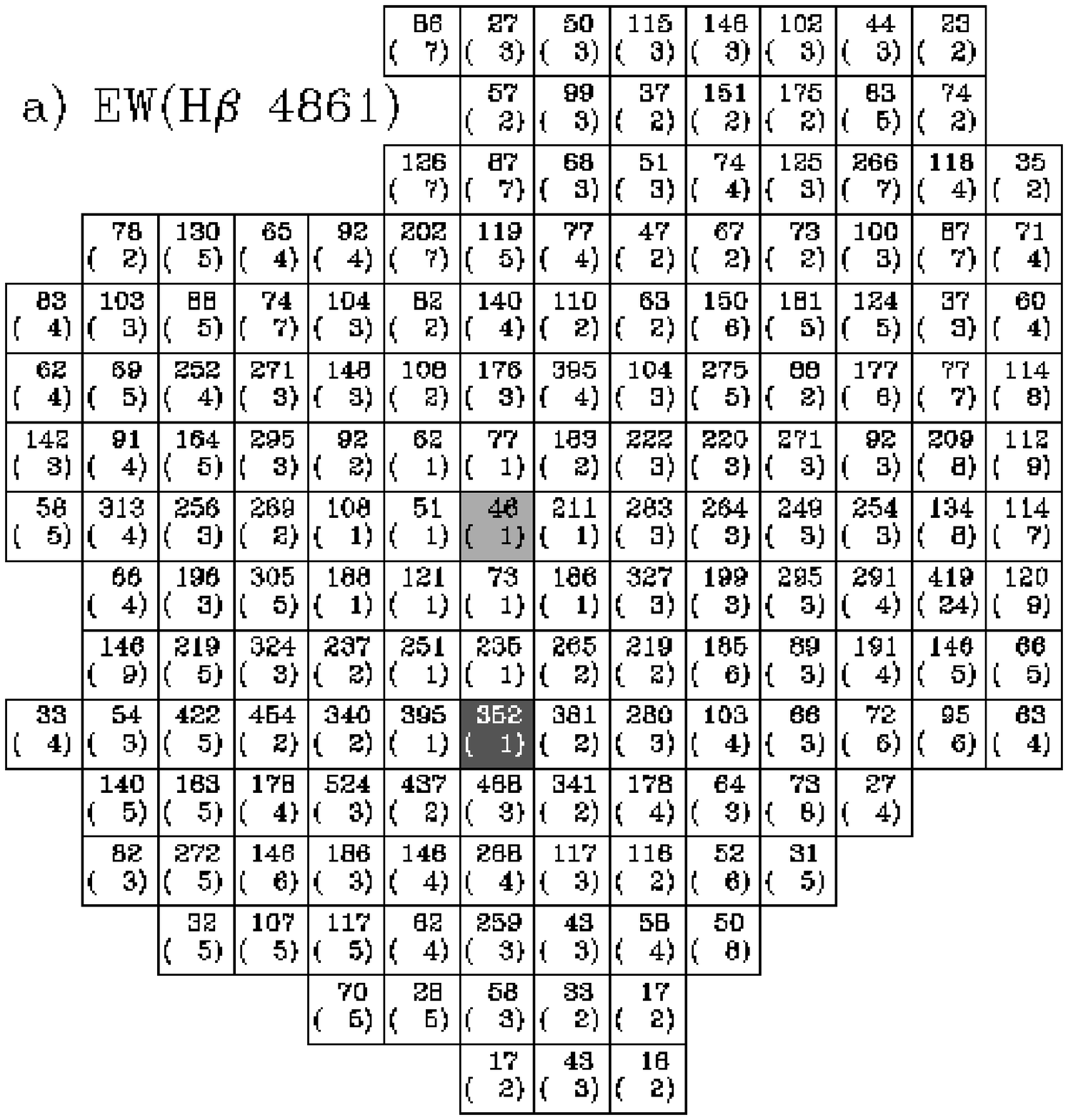,angle=0,width=6.0cm}
\hspace*{1.0cm}\psfig{figure=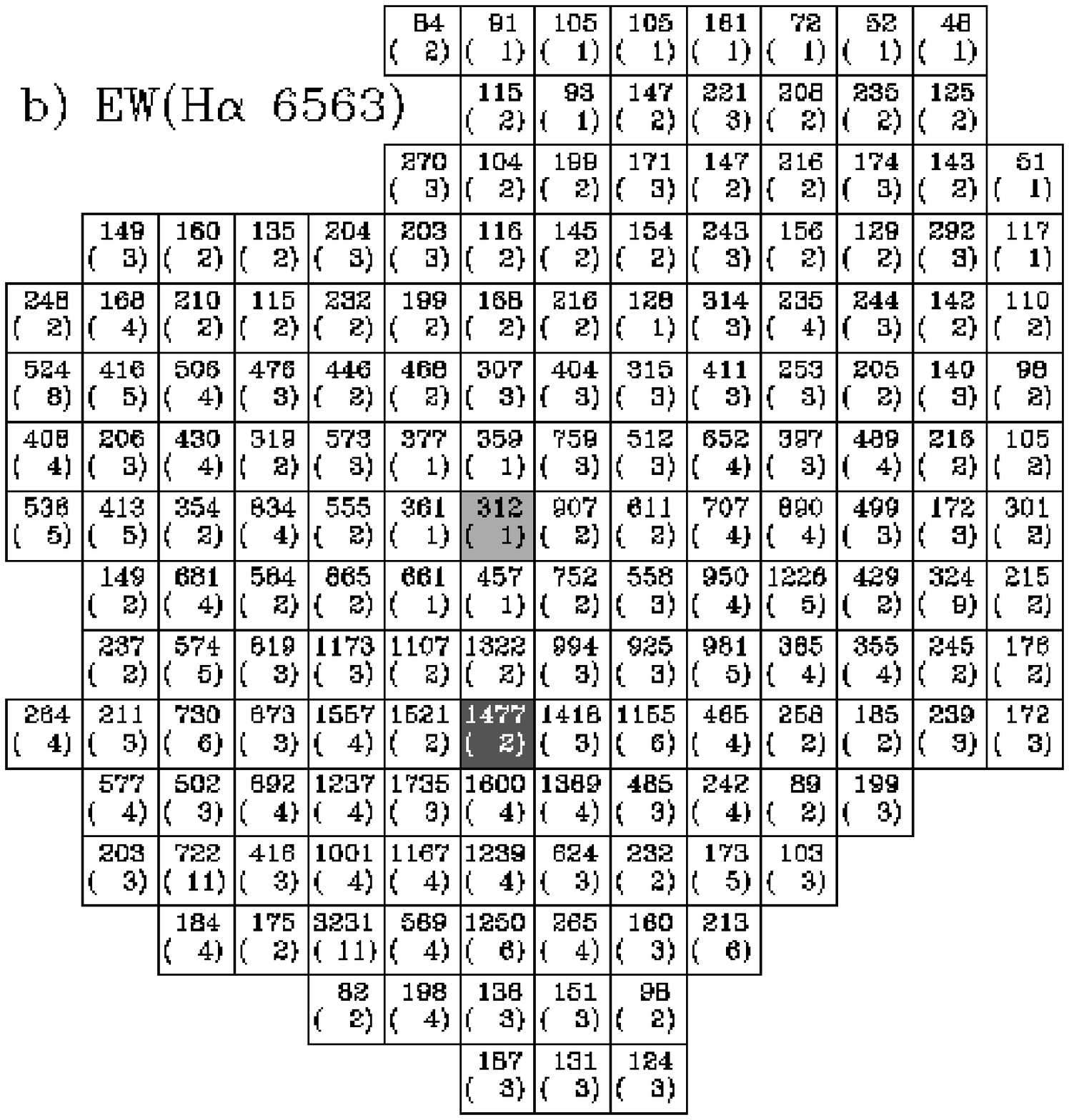,angle=0,width=6.0cm,clip=}
\caption{Distribution of the H$\beta$ $\lambda$4861\AA\ equivalent width (a)
and of the H$\alpha$ $\lambda$6563\AA\ equivalent width (b) in \AA, with
errors indicated in parentheses. Each square
region corresponds to 0\farcs52$\times$0\farcs52. Only regions 
in which both EW(H$\beta$) and EW(H$\alpha$) could be measured are shown.
Dark squares indicate the locations
of clusters 1+2 with the maximum flux of the H$\alpha$ emission line, 
grey squares the locations of clusters 4+5.
}
\label{fig5}
\end{figure*}

\section{Morphology in Continuum and Emission Lines \label{sec:morph}}

One of the advantages of the SBS 0335--052E panoramic observations with 
GIRAFFE/ARGUS is that the spectra of each region with an angular size of
0\farcs52$\times$0\farcs52 within an aperture 11\farcs4$\times$7\farcs3 
were obtained. This allows us to study the morphology of the galaxy
in the continuum and individual emission lines and to construct the model of
its H {\sc ii} region.

The central part of SBS 0335--052E containing the brightest clusters has an
angular size $\la$ 2\arcsec\ (Fig. \ref{fig1}) which is only $\la$ 4 times
larger than the angular size of 0\farcs52 of each ARGUS lens. Therefore, for 
better viewing we rebinned the images, splitting 
each pixel in 9 from 0\farcs52 pixel sizes to 0\farcs17
pixel sizes linearly interpolating flux values in adjacent 0\farcs52 pixels.
In Fig. \ref{fig4} are shown the rebinned images of SBS 0335--052E in the 
continuum near H$\beta$ $\lambda$4861\AA\ (a), and in the emission lines 
[O {\sc ii}] $\lambda$3727\AA\ (b), [O {\sc iii}] $\lambda$4363\AA\ (c),
He {\sc ii} $\lambda$4686\AA\ (d), [Ar {\sc iv}] $\lambda$4740\AA\ (e), 
H$\beta$ $\lambda$4861\AA\ (f), [O {\sc iii}] $\lambda$5007\AA\ (g), 
He {\sc i} $\lambda$5876\AA\ (h) and H$\alpha$ $\lambda$6563\AA\ (i). 
In all panels white crosses denote the pixel with the largest flux of 
the H$\alpha$ $\lambda$6563\AA\ emission line which is coincident with the
location of clusters 1+2 in Fig. \ref{fig1}.

The image in the continuum (Fig. \ref{fig4}a) with labelled clusters resembles
well the HST UV image (Fig. \ref{fig1}) despite the much lower angular 
resolution determined by a seeing of $\sim$1\arcsec\ (Table \ref{tab1}). 
Several clusters are seen. However, the angular resolution in 
the GIRAFFE data is not enough to separate clusters 1 and 2, and 4 and 5.

The images in all emission lines (except for the He {\sc ii} $\lambda$4686\AA\ 
line) are
very similar. They show very bright emission in the region of clusters
1, 2 and much fainter emission in the direction of other clusters. 
Furthermore, the equivalent widths of H$\beta$ $\lambda$4861\AA\ and 
H$\alpha$ $\lambda$6563\AA\ emission lines in the clusters 1 and 2 (dark
squares in Fig. \ref{fig5}) and in the regions around these clusters are high.
These facts suggest that clusters 1 and 2 are young, with an age 3--4 Myr,
and contain numerous hot and massive ionising main-sequence stars. 
It is likely, that clusters 7 and probably 3 are also young because 
EW(H$\beta$) and EW(H$\alpha$) are high.
However, the number of ionising massive stars in those clusters is much lower 
than in clusters 1 and 2.
On the other hand, clusters 4, 5 and 6 are probably more evolved as 
evidenced by their lower equivalent widths of H$\beta$ $\lambda$4861\AA\ and 
H$\alpha$ $\lambda$6563\AA\ emission lines. 
In particular, the H$\beta$ equivalent width
EW(H$\beta$) = 46\AA\ for clusters 4, 5 (grey square in 
Fig. \ref{fig5}) and 87\AA\ for the square region delineated by the thick
dashed line correspond to an age of $\sim$ 6 -- 8 Myr and $\sim$ 5 Myr adopting
heavy element mass fraction $Z$ = 0.001 for stars \citep{SV98}. Age for
clusters 4,5 is larger if lower $Z$ = 0.0004 corresponding to the metallicity 
of the ionised gas is adopted.
The larger age of clusters 4 and 5 is supported by their weak P$\alpha$
emission as compared to that of clusters 1 and 2 \citep{T06}.
The different age of clusters 1+2
and 4+5 could in principle explain why the brightness of clusters 4 and 5
in the UV range is greater than that of clusters 1 and 2 (Fig. \ref{fig1}),
why their brightness is comparable in the optical continuum (Fig. \ref{fig4}a)
and why clusters 4 and 5 are fainter in the NIR \citep{T06}. This is because
the relative contribution of the ionised gas emission is increased from the UV
to the NIR. The effect is stronger for clusters 1 and 2 because of the higher
EW(H$\beta$) and hence of higher contribution of the ionised gas to the
total emission. In addition, the interstellar extinction may play role if it
is higher for clusters 1 and 2.

\begin{figure}
\hspace*{1.0cm}\psfig{figure=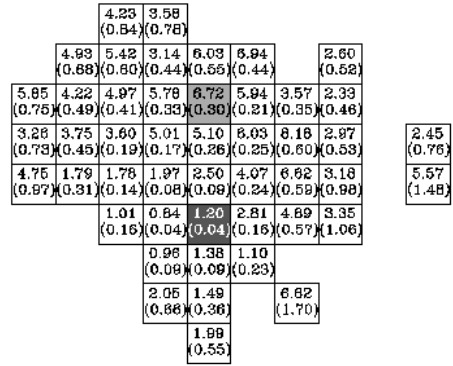,angle=0,width=6.0cm,clip=}
\caption{He {\sc ii} $\lambda$4686/H$\beta$ relative intensity distribution 
(in percent). Only regions with a strong enough He {\sc ii} $\lambda$4686 
emission line are shown. The dark square is the location
of clusters 1+2, while the grey square is the location of clusters 4+5.
}
\label{fig6}
\end{figure}

The morphology of SBS 0335--052E in the He {\sc ii} $\lambda$4686\AA\ emission line
(Fig. \ref{fig4}d) significantly differs from that in other emission lines.
The emission of this line in the direction of the clusters 3 and 4+5 
is stronger than that in the direction of the clusters 1 and 2.
This offset of the He {\sc ii} $\lambda$4686\AA\ emission 
line relative to other nebular
lines was noted earlier by \citet{I97b} and \citet{I01b}. Thus, it is evident
that the hard ionising radiation responsible for the 
He {\sc ii} $\lambda$4686\AA\ emission is not connected with the young 
main-sequence stars, but rather related to the post-main-sequence stars or
their remnants. This effect is more clearly seen in Fig. \ref{fig6} where we
show the distribution of the relative flux He {\sc ii} $\lambda$4686/H$\beta$.
In the direction on the clusters 1 and 2 the relative
He {\sc ii} $\lambda$4686/H$\beta$ flux is $\sim$ 1--2\% while in
north-west
regions it increases to $\sim$ 6 -- 7\%.
Such high relative fluxes of
the He {\sc ii} $\lambda$4686 emission line is difficult to
explain by the ionising stellar radiation \citep[e.g.]{I04,TI05}.
Although a small number of Wolf--Rayet stars are found in cluster 3 
\citep[see ][ and this paper]{P06}, 
other mechanisms such as radiative shocks need to be invoked.

\section{Kinematics of the Ionised Gas \label{sec:vel}}

The second advantage of the present GIRAFFE spectra for SBS 0335--052E is that
they are obtained with sufficient enough spectral resolution. Therefore, the
panoramic spectroscopic data can be used to study the kinematics of the
H {\sc ii} regions in this galaxy.

\begin{figure*}
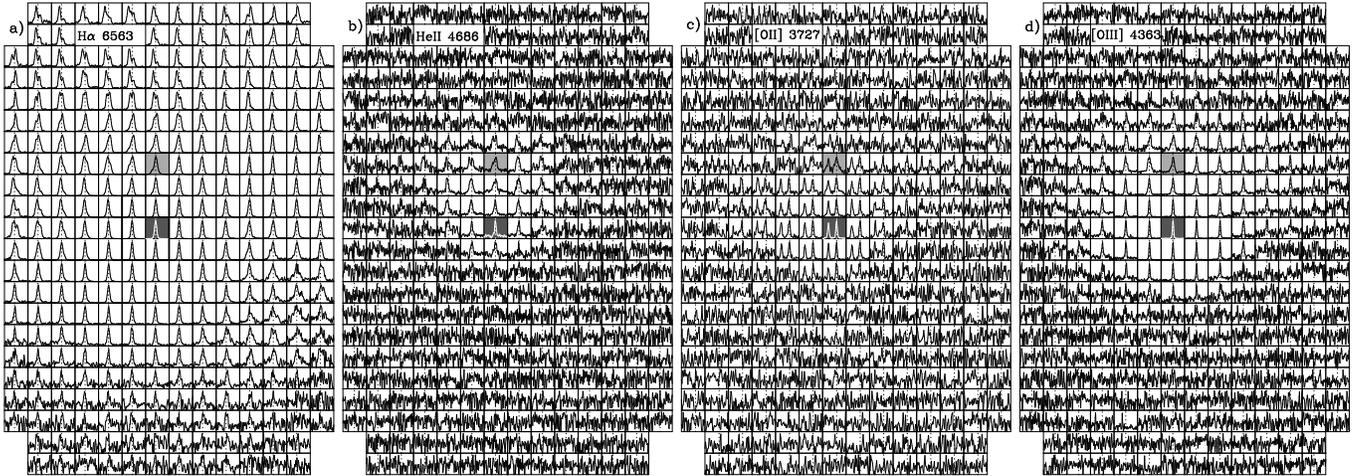

\hspace*{0.0cm}\psfig{figure=whaprofile.ps,angle=0,width=4.4cm}
\hspace*{0.1cm}\psfig{figure=wheii4686profile.ps,angle=0,width=4.4cm}
\hspace*{0.1cm}\psfig{figure=woii3727profile.ps,angle=0,width=4.4cm}
\hspace*{0.1cm}\psfig{figure=woiii4363profile.ps,angle=0,width=4.4cm,clip=}
\caption{Profiles of the emission lines: (a) H$\alpha$ $\lambda$6563\AA, (b) 
He {\sc ii} $\lambda$4686\AA, (c) [O {\sc ii}] $\lambda$3726,3729\AA\ and
(d) [O {\sc iii}] $\lambda$4363\AA. All 0\farcs52$\times$0\farcs52
regions are shown. Dark squares mark the location
of clusters 1+2, grey squares the location of clusters 4+5. Thick solid 
and thick dashed lines in (a) delineate respectively the brightest rectangular 
region and the second brightest square region for which the spectra are
shown in Fig. \ref{fig2} and \ref{fig3}.
}
\label{fig7}
\end{figure*}

In Fig. \ref{fig7} we show the profiles of the 
H$\alpha$ $\lambda$6563\AA\ (a), He {\sc ii} $\lambda$4686\AA\ (b), 
[O {\sc ii}] $\lambda$3726, 3729\AA\ (c) and [O {\sc iii}] $\lambda$4363\AA\ 
(d) emission
lines in each pixel of the ARGUS array. Dotted lines show the wavelengths
of emission lines adopting the average redshift derived from the observed 
wavelengths of all strong emission lines in the spectrum of the 
brightest rectangular region delineated by a thick solid
line in Fig. \ref{fig7}a. H$\alpha$ $\lambda$6563\AA\ is the strongest
line in all ARGUS array spectra and thus allows to study the kinematical 
properties in the
low-intensity extended regions of SBS 0335--052E. Other lines originate
in different zones of the H {\sc ii} region: He {\sc ii} $\lambda$4686\AA\ 
is a characteristic of the highest ionisation zone, 
[O {\sc ii}] $\lambda$3726, 3729\AA\ are 
characteristics of the lowest ionisation zone, 
and [O {\sc iii}] $\lambda$4363\AA\ 
emission corresponds to the intermediate zone which is overlapped with the
highest- and lowest-ionisation zones.

\begin{figure}[t]
\hspace*{1.5cm}\psfig{figure=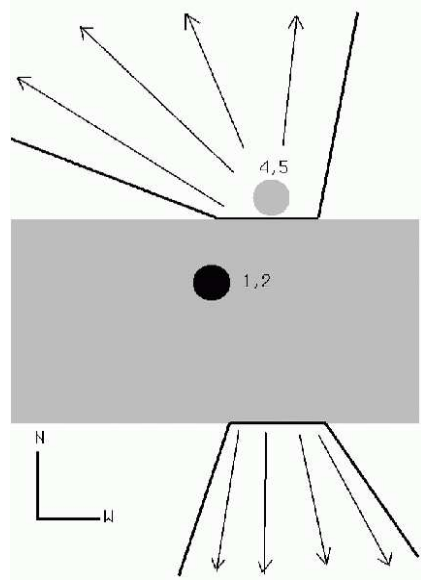,angle=0,width=5.5cm}
\caption{Schematic H$\alpha$ kinematic model of SBS 0335--052E. Clusters 
1,2 and 4,5 are shown by dark and grey circles respectively. The region with narrow H$\alpha$
profiles is shown by a grey rectangle. Its orientation is close to the 
orientation of the H {\sc i} cloud discussed by \citet{P01}. Two regions with
the ionised gas outflow are located to the north and south from the region
with narrow H$\alpha$ emission line.
}
\label{fig8}
\end{figure}

H$\alpha$ emission is seen almost in the whole region observed with GIRAFFE
(Fig. \ref{fig7}a). The total aperture 11\farcs4$\times$7\farcs3 of ARGUS
corresponds to linear size $\sim$3.1 kpc$\times$1.8 kpc adopting the distance
to SBS 0335--052E of 54.3 Mpc \citep{I97b}. Thus,
the observed region is only a part of a much larger H {\sc ii} region
with a size of $\sim$ 6--8 kpc detected by \citet{I01b}. 
In the brightest 
central region and in the slice oriented west-east the 
H$\alpha$ $\lambda$6563 line is narrow and no systematic offset of 
the line profile from the dotted line is seen. Thus, no evidence is present for
the rotation of the observed part of the galaxy, since the 
west-east orientation of the region
with narrow profiles is close to that for the disk-like H {\sc i} cloud
seen edge-on \citep{P01}. On the other hand, in the north-south direction
the H$\alpha$ profiles are much broader and with more complex structure,
suggesting an outflow of ionised gas in the direction perpendicular
to the H {\sc i} disk. 
The schematic H$\alpha$ kinematic model is shown in Fig. \ref{fig8}
where the grey rectangular region is the region with narrow H$\alpha$
profiles which is oriented approximately along the H {\sc i} cloud
detected by \citet{P01}. Two regions with the ionised gas outflow are
shown to the north and south of the region with the narrow H$\alpha$ line.
The double-peaked H$\alpha$ profiles in the northern
and southern parts of Fig. \ref{fig7}a suggest the presence of expanding
shells of ionised hydrogen with radial components of velocities of
$\sim 50$ km s$^{-1}$. This finding is consistent with the presence of
a shell of ionised gas at an angular distance of $\sim$ 5\arcsec\ 
to the north from the cluster 1 \citep{T97}.
The width of H$\alpha$ in clusters 4, 5 
(grey square in Fig. \ref{fig7}a) is $\sim$ 2 times larger than that in 
clusters 1 and 2 (dark square in Fig. \ref{fig7}a)
implying higher dynamic activity of the interstellar medium in the former,
older clusters. This higher activity is probably due to supernova
explosions.

\begin{figure*}[t]
\hspace*{-0.0cm}\psfig{figure=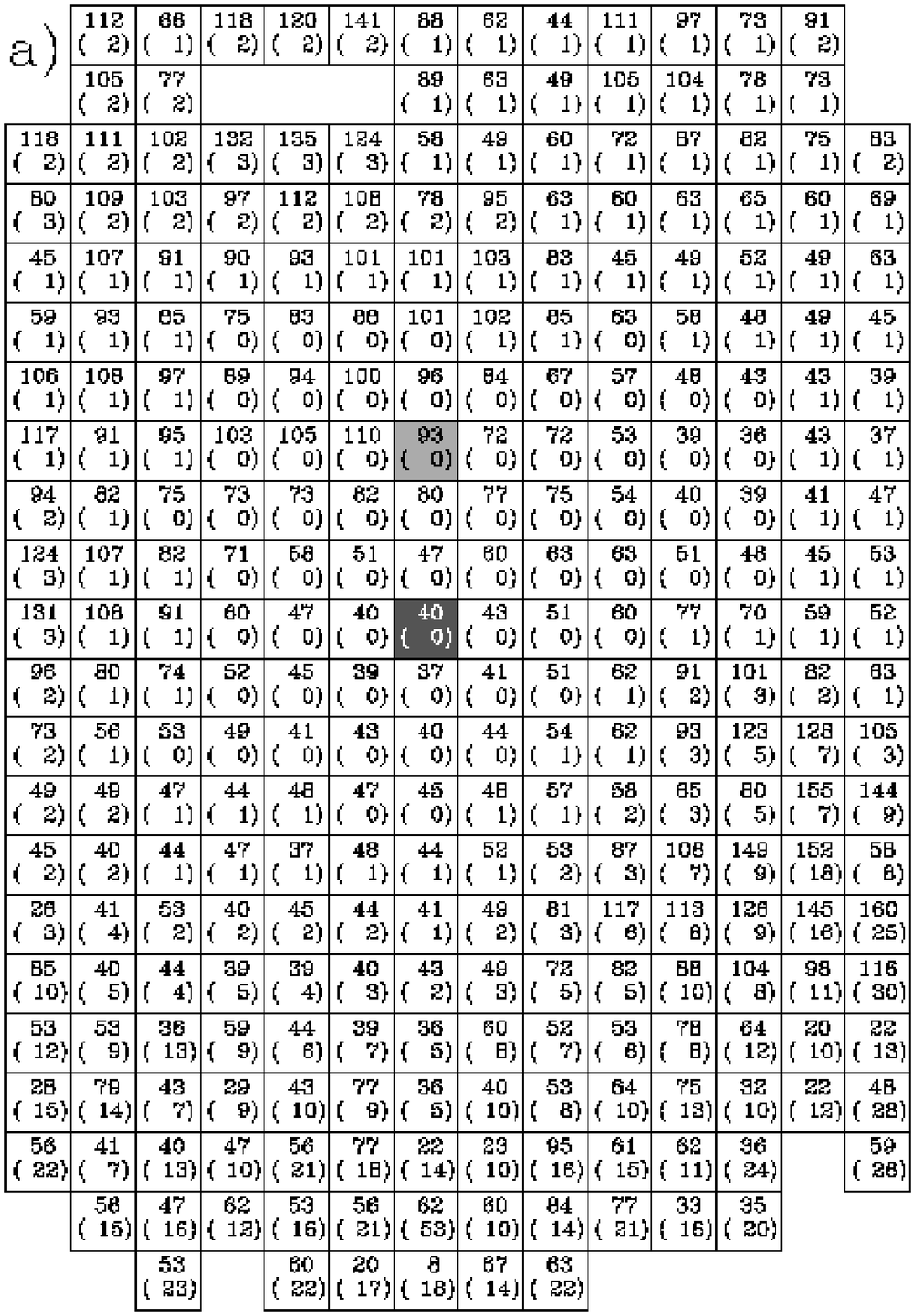,angle=0,width=5.0cm}
\hspace*{0.1cm}\psfig{figure=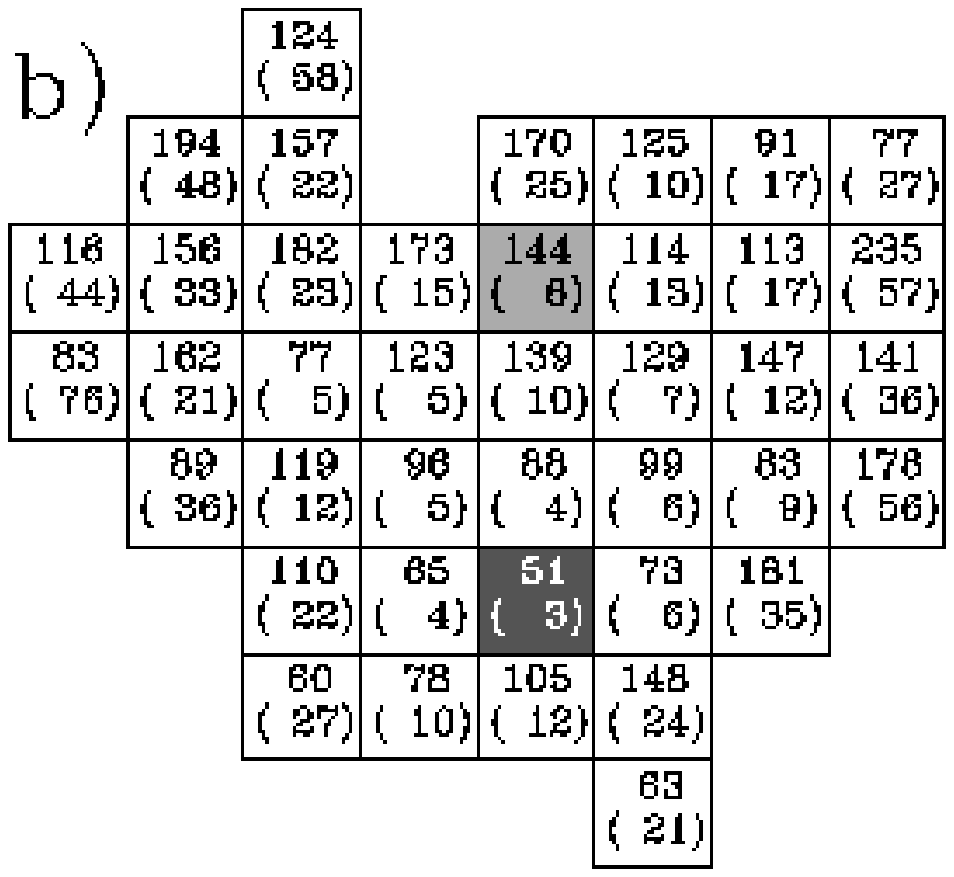,angle=0,width=3.0cm}
\hspace*{0.1cm}\psfig{figure=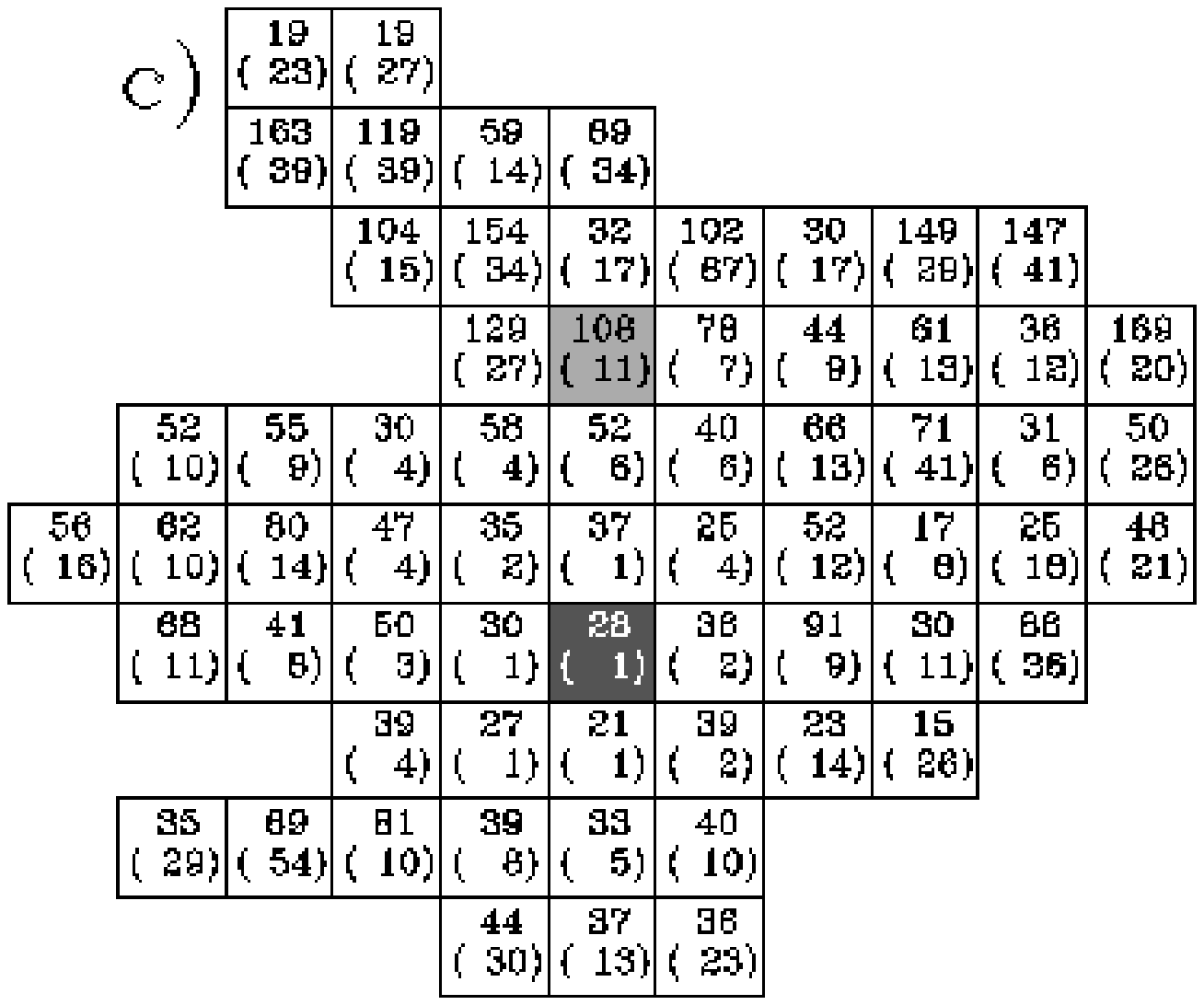,angle=0,width=4.0cm}
\hspace*{0.1cm}\psfig{figure=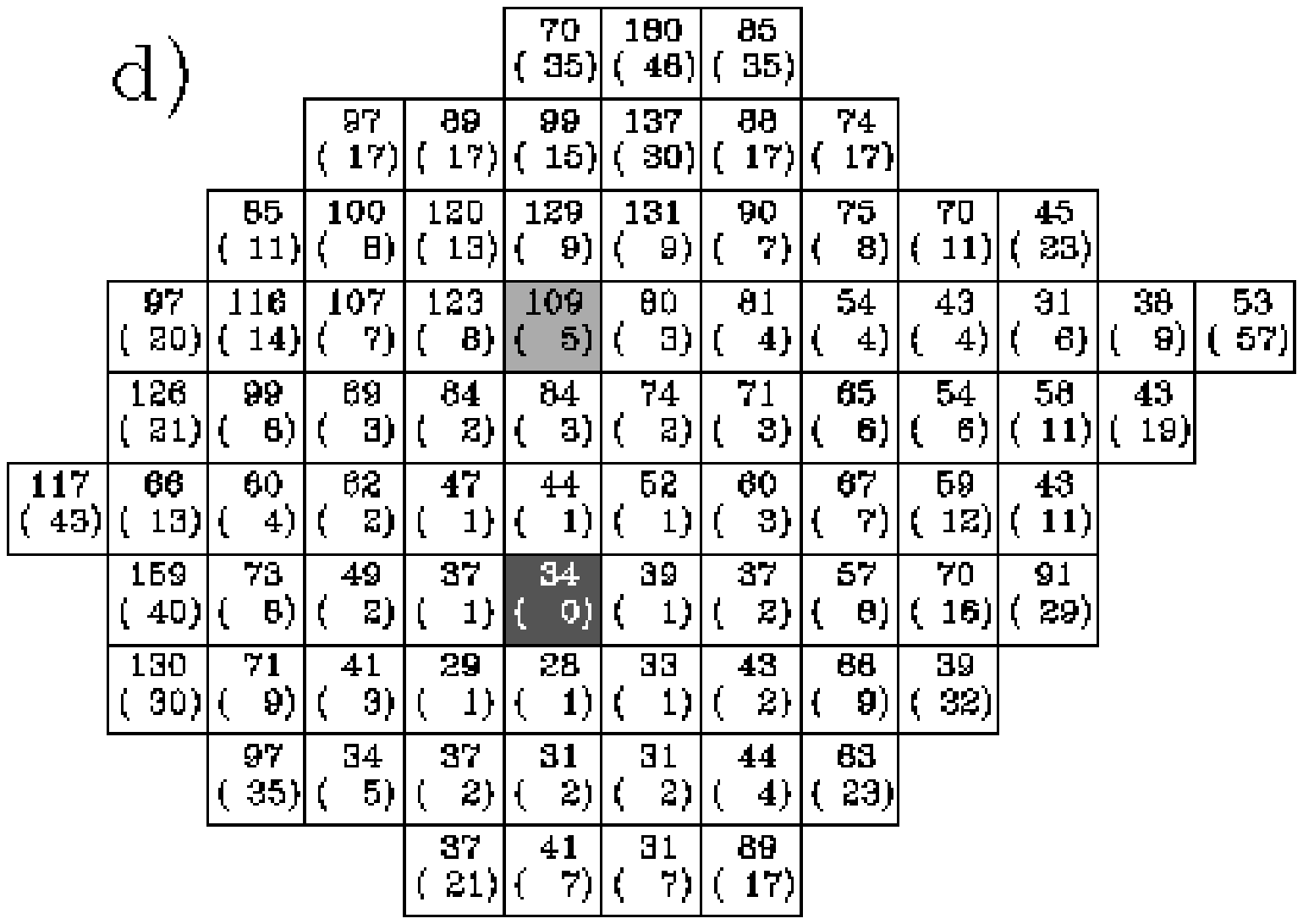,angle=0,width=5.0cm,clip=}
\caption{Macroscopic/turbulent velocity dispersion at the FWHM (in km s$^{-1}$) derived
from the (a) H$\alpha$ $\lambda$6563\AA, (b) He {\sc ii} $\lambda$4686\AA,
(c) [O {\sc ii}] $\lambda$3729\AA\ and (d) [O {\sc iii}] $\lambda$4363\AA\ 
emission lines. The errors of the velocity dispersion are shown in parentheses.
Only regions with the detected lines are shown. 
The dark square shows the location of
clusters 1+2 with a maximum flux of H$\alpha$ 6563\AA\ 
emission line. The grey square shows the location of clusters 4+5.
}
\label{fig9}
\end{figure*}

All observed profiles were fit with a single gaussian profile.
In Fig. \ref{fig9}a we show FWHMs of the H$\alpha$ emission line in
km s$^{-1}$ due to the macroscopic motion only. 
Errors are given in parentheses.
The widths are obtained after correction for the instrumental profiles and 
for the thermal motion, following the formula
\begin{equation}
{\rm FWHM}^2_{tur}={\rm FWHM}^2_{obs}-{\rm FWHM}^2_i-{\rm FWHM}^2_{th},
\end{equation}
where FWHM$_{tur}$, FWHM$_{obs}$, FWHM$_{i}$ and FWHM$_{th}$ are the
width of the macroscopic motion, the observed width, the 
width of the instrumental
profile and the width of the thermal motion, respectively.
The FWHM$_{i}$ is obtained from the profiles of night sky emission lines
and it is $\sim$ 25 km s$^{-1}$. The width of the thermal profile of 
30 km s$^{-1}$ is derived for the electron temperature $T_e$ = 20000K and
it is applied to the H$\alpha$ $\lambda$6563\AA, He {\sc ii} $\lambda$4686\AA\ 
and [O {\sc iii}] $\lambda$4363\AA\ emission lines, while the width of
26 km s$^{-1}$ is obtained for $T_e$ = 15000K and it is applied to
the [O {\sc ii}] $\lambda$3729\AA\ emission line.

In the brightest part of SBS 0335--052E the FWHM(H$\alpha$) in and around
the clusters 1 and 2 is $\sim$ 40 km s$^{-1}$. This value is lower than
the FWHM(H$\alpha$) of 83 km s$^{-1}$ and 62 km s$^{-1}$ derived by 
\citet{Pe97} for the NW and SE components of I Zw 18. Apparently, the 
dispersion of macroscopic motion is strongly dependent on the age of starburst
and it is larger in the evolved starbursts where the SNe activity is higher.
Indeed, the difference between clusters 1 and 2 in SBS 0335--052E and the
NW and SE components of I Zw 18 is that the clusters 1 and 2 are
younger as the equivalent width of the H$\beta$ emission line there,
EW(H$\beta$) $\sim$ 300 -- 400\AA\ (Fig. \ref{fig5}a), is much larger than
the EW(H$\beta$) $\la$ 100\AA\ in the NW and SE components of I Zw 18.
It is probable that SNe have not yet appeared or their number is small
in clusters 1 and 2.

On the other hand, the macroscopic motion in and around the older 
clusters 4 and 5 is significantly larger with  
FWHM(H$\alpha$) $\sim$ 100 km s$^{-1}$, likely due to a high SN activity.
Similar FWHM values and a similar spatial behaviour are found for the 
[O {\sc ii}] $\lambda$3729\AA\ and [O {\sc iii}] $\lambda$4363\AA\ 
emission lines. (Figs. \ref{fig9}c and \ref{fig9}d). 
The tendency of higher FWHM in more evolved clusters is also retained for 
the He {\sc ii} $\lambda$4686\AA\ emission line (Fig. \ref{fig9}b). 
However, the macroscopic
velocity in the regions of He {\sc ii} emission is significantly larger than
that in the emission regions of other lines. This difference may be an
additional indication that the source of hard radiation is connected with fast
radiative shocks.

\section{Physical Conditions and Element Abundances \label{sec:abund}}

The large aperture (11\farcs4$\times$7\farcs3), high enough spectral resolution and
large wavelength coverage of the ARGUS observations allow the detailed study
of physical conditions (electron temperature and electron number density) and
element abundances in the H {\sc ii} region. 
However, there are some limitations to this study. First,
observations in different wavelength ranges have been done not in single
but in separate exposures during several nights (see Table \ref{tab1}). 
Therefore, due to the varying weather conditions (seeing, transparency),
effects of the atmospheric refraction and non-perfect pointing during
different exposures the spectra in small apertures such as in a single pixel 
of 0\farcs52$\times$0\farcs52 and in different wavelength ranges are not quite
well adjusted since they represent slightly different regions of the galaxy.
These effects tend to be lower with increasing aperture. Therefore, depending
on the adopted aperture, we will follow different approaches in the 
determination of element abundances. We consider the element abundance
determination from the spectra obtained in apertures with three different 
sizes. First, for the 
spectra obtained within the largest available  aperture of 
11\farcs4$\times$7\farcs3 there is no need to adjust the different wavelength
ranges. This allows to correct consistently the spectra for interstellar 
extinction using the observed decrement of hydrogen Balmer lines and then
derive element abundances. Second, we consider the spectrum of the brightest
region of SBS 0335--052E obtained within an aperture 1\farcs56$\times$1\farcs04
(the rectangular region delineated by a thick solid line in Fig. \ref{fig7}a). 
This spectrum is shown in
Fig. \ref{fig2}. Since the aperture for this spectrum is relatively small, some
adjustment of adjacent wavelength ranges is needed. However, thanks to the high
brightness of this region many bright emission lines are seen in its spectrum.
Therefore we use the same brightest emission lines in the overlapping 
wavelength ranges (where this is possible) to scale the spectra in the 
adjacent wavelength ranges. These lines are H7 $\lambda$3970\AA, 
He {\sc i} $\lambda$4471\AA, [O {\sc iii}] $\lambda$4959\AA, 
[O {\sc i}] $\lambda$6300\AA\ and He {\sc i} $\lambda$7065\AA. 
In the remaining two overlapping wavelength
ranges $\lambda$5650--5750 and $\lambda$8100--8200
where no strong emission lines are seen we used the continuum
levels to adjust the adjacent spectra. Thus, in the spectrum of the
brightest region the determination
of the interstellar extinction is still possible, which was used to correct the 
spectra. Third, in the case of smallest apertures
of 0\farcs52$\times$0\farcs52, in general the signal-to-noise of the spectra is not
enough to use the same emission lines in the overlapping wavelength ranges.
Therefore, for these apertures we adjust spectra in different wavelength ranges
assuming that the ratios of hydrogen Balmer lines correspond to the theoretical
values at the electron temperature of $T_e$ = 20000K. Hence, no determination 
of the interstellar extinction is possible
for the smallest apertures and not all wavelength ranges could be adjusted.
Fortunately, it is possible to adjust wavelength ranges containing the [O {\sc ii}]
$\lambda$3726, 3729\AA, [O {\sc iii}] $\lambda$4363, 4959, 5007\AA,
[S {\sc ii}] $\lambda$6717, 6731\AA\ emission lines,
therefore at least the determination of the electron temperature 
$T_e$(O {\sc iii}), the electron number density $N_e$(S {\sc ii}) and the
oxygen abundance is possible.

To derive $T_e$, $N_e$ and heavy element abundances we follow the prescriptions
by \citet{I05b}. Namely, where possible, the coefficient of interstellar
extinction $C$(H$\beta$) and the equivalent width of absorption hydrogen lines
EW$_{abs}$ are derived from the observed hydrogen Balmer decrement. In this
procedure we assume that EW$_{abs}$ is the same for all hydrogen lines. Then
the fluxes of emission lines were corrected for interstellar extinction
and underlying stellar absorption (where this is possible).

We adopt the three zone model of the H {\sc ii} region. The electron 
temperature $T_e$(O {\sc iii}) in the high-ionisation zone is derived from
the [O {\sc iii}] $\lambda$4363/($\lambda$4959+$\lambda$5007) flux ratio.
This temperature is used to derive abundances of ions O$^{2+}$ and Ne$^{2+}$.
Since He {\sc ii} $\lambda$4686\AA\ emission is present in the SBS 0335--052E 
spectrum, the O$^{3+}$ abundance is derived following \citet{I05b} and adopting
$T_e$(O {\sc iii}). Since the O$^{3+}$ abundance is significantly lower than
the O$^{2+}$ abundance, the uncertainties in the temperature for the zone
where the O$^{3+}$ ion is present introduce only a small uncertainty in the total
oxygen abundance. Some other emission lines of high-ionisation
ions Ar$^{3+}$, Cl$^{3+}$, Fe$^{3+}$ -- Fe$^{6+}$ are seen in the spectrum of 
SBS 0335--052E in Fig. \ref{fig2}. In general these ions are present in the
inner part of the H {\sc ii} region with a temperature higher than 
$T_e$(O {\sc iii}).

However, since there is no temperature 
constraint from observations for
these ions and atomic data are not well known for some of them,
we decided not to use these ions for the abundance determination.
For the intermediate-ionisation zone we adopt the electron 
temperature $T_e$(S {\sc iii}) which was derived using the relation
between $T_e$(S {\sc iii}) and $T_e$(O {\sc iii}) from \citet{I05b}. The
temperature $T_e$(S {\sc iii}) is used to derive the abundances of ions
S$^{2+}$, Ar$^{2+}$ and Cl$^{2+}$. Finally, for the low-ionisation zone
we adopt the electron 
temperature $T_e$(O {\sc ii}) which was derived using the relation
between $T_e$(O {\sc ii}) and $T_e$(O {\sc iii}) from \citet{I05b}. This
temperature is used to derive the abundances of ions O$^{+}$, N$^+$, S$^+$ and
Fe$^{2+}$. Emission lines of some other low-ionisation ions and neutral 
species are present in the low-ionisation zone, namely, Fe$^+$, O$^0$, N$^0$.
We decided not to use them for the abundance determination since they reside
both in the neutral and ionised gas and their abundances in the H {\sc ii}
region are subject to large uncertainties.

The electron number density $N_e$ is derived using the 
[O {\sc ii}] $\lambda$3726/$\lambda$3729, 
[Ar {\sc iv}] $\lambda$4711/$\lambda$4740 and
[S {\sc ii}] $\lambda$6717/$\lambda$6731 emission line fluxes.

The total heavy element abundances are obtained with the use of ionisation
correction factors $ICF$s from \citet{I05b} for every element. $ICF$s take 
into account the abundances of ions in unseen stages of ionisation.

\subsection{Heavy Element Abundances in the Brightest Region}

We first consider the chemical composition in the brightest part of 
SBS 0335--052E. Observed emission line fluxes 
relative to the H$\beta$ flux, 
$F$($\lambda$)/$F$(H$\beta$), emission line fluxes relative to the 
H$\beta$ flux, corrected for interstellar extinction
and underlying stellar absorption, $I$($\lambda$)/$I$(H$\beta$), and the
equivalent widths of emission lines 
EW($\lambda$) are shown in Table \ref{tab2}.
The extinction coefficient $C$(H$\beta$), the equivalent width EW$_{abs}$
of the absorption hydrogen Balmer lines and the observed flux $F$(H$\beta$)
of the H$\beta$ emission line are shown at the end of Table 
\ref{tab2}. 
Electron temperatures, electron number densities, ionic and total heavy
element abundances for the brightest region are shown in the second
column of Table \ref{tab3}. In general, the derived parameters are
consistent with previous determinations e.g. by \citet{I97b}, \citet{I99},
\citet{TI05}. In particular, the electron temperature $T_e$(O {\sc iii}) 
in all measurements is high and is close to 20000K. It was found in previous
studies that the H {\sc ii} region in SBS 0335--052E is relatively dense.
We confirm this finding. The electron number density, which we derive
from the [S {\sc ii}] $\lambda$6717/$\lambda$6731 flux ratio
in the brightest region, is $\sim$ 150 cm$^{-3}$. A similar value is obtained
from the [O {\sc ii}] $\lambda$3726/$\lambda$3729 flux ratio. On the other
hand, the electron number density derived from the [Ar {\sc iv}] 
$\lambda$4711/$\lambda$4740 flux ratio, using the data from \citet{A84}, 
is much larger, $\sim$ 6000 cm$^{-3}$. Thus, it appears that the 
high-ionisation regions are much denser than the low-ionisation regions.



\begin{table}
\caption{Heavy Element Abundances \label{tab3}}
\begin{tabular}{lcc} \hline
Property       &Bright&Total \\ \hline
$T_e$(O {\sc iii}), K           &20120$\pm$240  &20360$\pm$480 \\
$T_e$(O {\sc ii}), K            &15620$\pm$170  &15630$\pm$340 \\
$T_e$(S {\sc iii}), K          &18750$\pm$200  &18910$\pm$400 \\
$N_e$(O {\sc ii}), cm$^{-3}$    &120$\pm$50     &150$\pm$80    \\
$N_e$(Ar {\sc iv}), cm$^{-3}$   &6000$\pm$2000   &    ...    \\
$N_e$(S {\sc ii}), cm$^{-3}$    &146$\pm$41     &213$\pm$74    \\ \\
O$^+$/H$^+$, ($\times$10$^5$)   &0.164$\pm$0.005&0.211$\pm$0.013  \\
O$^{2+}$/H$^+$, ($\times$10$^5$)&1.838$\pm$0.054&1.664$\pm$0.090  \\
O$^{3+}$/H$^+$, ($\times$10$^5$)&0.032$\pm$0.001&0.081$\pm$0.006  \\
O/H, ($\times$10$^5$)           &2.034$\pm$0.054&1.956$\pm$0.091  \\
12+log O/H                      &7.31$\pm$0.01  &7.29$\pm$0.02 \\ \\
N$^+$/H$^+$, ($\times$10$^7$)   &0.779$\pm$0.019&0.673$\pm$0.033  \\
$ICF$(N)                        &     11.25     &      8.53     \\
N/H, ($\times$10$^7$)           &8.765$\pm$0.235&5.741$\pm$0.304  \\
log N/O                         &--1.37$\pm$0.02&--1.53$\pm$0.03  \\ \\
Ne$^{2+}$/H$^+$, ($\times$10$^6$)&2.977$\pm$0.092&2.941$\pm$0.199 \\
$ICF$(Ne)                        &      1.04     &      1.06     \\
Ne/H, ($\times$10$^6$)           &3.099$\pm$0.102&3.126$\pm$0.234  \\
log Ne/O                         &--0.82$\pm$0.02&--0.80$\pm$0.04  \\ \\
S$^{+}$/H$^+$, ($\times$10$^7$) &0.314$\pm$0.006 &0.348$\pm$0.013 \\
S$^{2+}$/H$^+$, ($\times$10$^7$)&1.539$\pm$0.052 &1.433$\pm$0.136 \\
$ICF$(S)                        &      2.36      &      1.93     \\
S/H, ($\times$10$^7$)           &4.363$\pm$0.122 &3.431$\pm$0.264 \\
log S/O                         &--1.67$\pm$0.02 &--1.76$\pm$0.04 \\ \\
Cl$^{2+}$/H$^+$, ($\times$10$^9$)&2.846$\pm$0.560&       ...      \\
$ICF$(Cl)                        &      1.48     &       ...      \\
Cl/H, ($\times$10$^9$)           &4.207$\pm$0.827&       ...      \\
log Cl/O                         &--3.68$\pm$0.09&       ...      \\ \\
Ar$^{2+}$/H$^+$, ($\times$10$^8$)&4.478$\pm$0.094&4.408$\pm$0.199  \\
$ICF$(Ar)                        &      1.49     &      1.34     \\
Ar/H, ($\times$10$^8$)           &6.680$\pm$0.287&5.900$\pm$0.812  \\
log Ar/O                         &--2.48$\pm$0.02&--2.52$\pm$0.06  \\ \\
Fe$^{2+}$/H$^+$, ($\times$10$^7$)&0.453$\pm$0.033&       ...      \\
$ICF$(Fe)                        &      17.02    &       ...       \\
Fe/H, ($\times$10$^7$)           &7.717$\pm$0.567&       ...      \\
log Fe/O                         &--1.42$\pm$0.03&       ...      \\
\hline
\end{tabular}
\end{table}


The oxygen abundance 12+log O/H = 7.31 $\pm$ 0.01 is in perfect
agreement with recent determinations by \citet{I97b}, \citet{I99} and
\citet{TI05}. The Ne/O, S/O, Cl/O and Ar/O abundance ratios are 
very close to the average values found by, e.g., \citet{I05b} for
the large sample of low-metallicity emission-line galaxies. On the
other hand, the N/O abundance ratio --1.37 appears higher than the mean
value of --1.5 to --1.6 for the most metal-deficient BCDs \citep{IT99,I05b}. 
Since only N$^+$ lines are observed in the optical spectrum of SBS 0335--052E
the total nitrogen abundance is derived as N/H = $ICF$(N)$\times$N$^+$/H$^+$,
where $ICF$(N) $\sim$ (O$^{3+}$+O$^{2+}$+O$^+$)/O$^+$. Inspection
of Table \ref{tab2} shows that the relative flux 
[O {\sc ii}] $\lambda$3727/H$\beta$ is 0.2, or 30\% lower than
that in some other observations of SBS 0335--052E \citep[e.g. ][]{I99,P06}
resulting in high $ICF$(N). The lower [O {\sc ii}] $\lambda$3727 flux in the bright
region (Table \ref{tab2}) could be due to observational uncertainties
(slightly different pointings of SBS 0335--052E during observations with 
different setups, effect of differential refraction, variable seeing, etc.).
Adopting a [O {\sc ii}] $\lambda$3727 flux $\sim$ 30\% higher will result
in log N/O $\sim$ --1.5, in much better agreement with other determinations.
Such increase of the [O {\sc ii}] $\lambda$3727 flux will also slightly
decrease by $\sim$ 0.1 dex the iron abundance, while the abundances of
other heavy elements will remain almost unchanged. In particular, the oxygen
abundance 12 + logO/H will be increased only by 0.01 dex. The
Fe/O abundance ratio is high and is typical for the extremely
metal-deficient BCDs \citep{I05b}. This fact suggests that the
depletion of iron onto dust in SBS 0335--052E is small.

\subsection{Heavy Element Abundances from the Integrated  Spectrum of
SBS 0335--052E}

We use panoramic VLT/GIRAFFE data to obtain the integrated spectrum of
SBS 0335--052E by summing every of 22$\times$14 spectra in the whole
aperture. The resulting spectrum is significantly more noisy as compared
to the spectrum of the brightest region because many spectra of
low-brightness regions were co-added to the spectrum of the brightest
region. However, the integrated spectrum is not subject to the observational
uncertainties which are much more important for the spectra obtained
with the smaller apertures (non-perfect pointing, variable seeing).
Additionally, it allows to obtain integrated characteristics such as
the luminosity of the galaxy in individual lines.

In Table \ref{tab4} are shown the measured absolute fluxes $F$($\lambda$)
of the emission lines, the absolute fluxes $I$($\lambda$) corrected for
the interstellar extinction and underlying stellar absorption, the respective
fluxes relative to the H$\beta$ $\lambda$4861 flux, 
$F$($\lambda$)/$F$(H$\beta$) and $I$($\lambda$)/$I$(H$\beta$),
the equivalent widths EW($\lambda$) of the emission lines, the interstellar
extinction coefficient $C$(H$\beta$) and equivalent width of hydrogen
Balmer absorption lines EW$_{abs}$. The absolute measured flux of
H$\alpha$ emission line 
$F$(H$\alpha$) = 3.45$\times$10$^{-13}$ erg s$^{-1}$ cm$^{-2}$ is consistent
with the value 3.23$\times$10$^{-13}$ erg s$^{-1}$ cm$^{-2}$ obtained by
\citet{P04}. The luminosities of the H$\beta$ and H$\alpha$ emission lines
corrected for interstellar extinction and underlying stellar absorption
are equal to $L$(H$\beta$) = 6.4$\times$10$^{40}$ erg s$^{-1}$ and
$L$(H$\alpha$) = 1.8$\times$10$^{41}$ erg s$^{-1}$, corresponding to the
equivalent number of O7{\sc v} stars $N$(O7{\sc v}) = 1.3$\times$10$^4$.
The major fraction of these stars ($\ga$ 90\%) is located in the two compact 
clusters 1 and 2 as it is evidenced by the high-resolution spatial
distribution of the P$\alpha$ emission in SBS 0335--052E obtained by
\citet{T06} from the HST observations. 
To the best of our knowledge, these two clusters (most likely, cluster 1)
are among the richest super-star clusters, hosting a very large number of O 
stars within a region of angular size $\la$ 0\farcs1 -- 0\farcs2, 
corresponding to a linear size $\la$ 25 -- 50 pc.

Electron temperatures, electron number densities $N_e$(S {\sc ii}) and
$N_e$(O {\sc ii}), ionic and
total element abundances derived from the integrated spectrum are shown in
the third column of Table \ref{tab3}. They are similar to the parameters
derived for the brightest region despite the fact that the statistical errors
for the parameters derived from the integrated spectrum are higher.
Note that the N/O abundance ratio derived
from the integrated spectrum is lower than that derived from the spectrum of
the brightest region and is consistent with the average values of N/O
obtained for most metal-deficient galaxies \citep{IT99,I05b}.

\begin{figure*}
\hspace*{0.0cm}\psfig{figure=tempdistr.ps,angle=0,width=5.0cm}
\hspace*{1.0cm}\psfig{figure=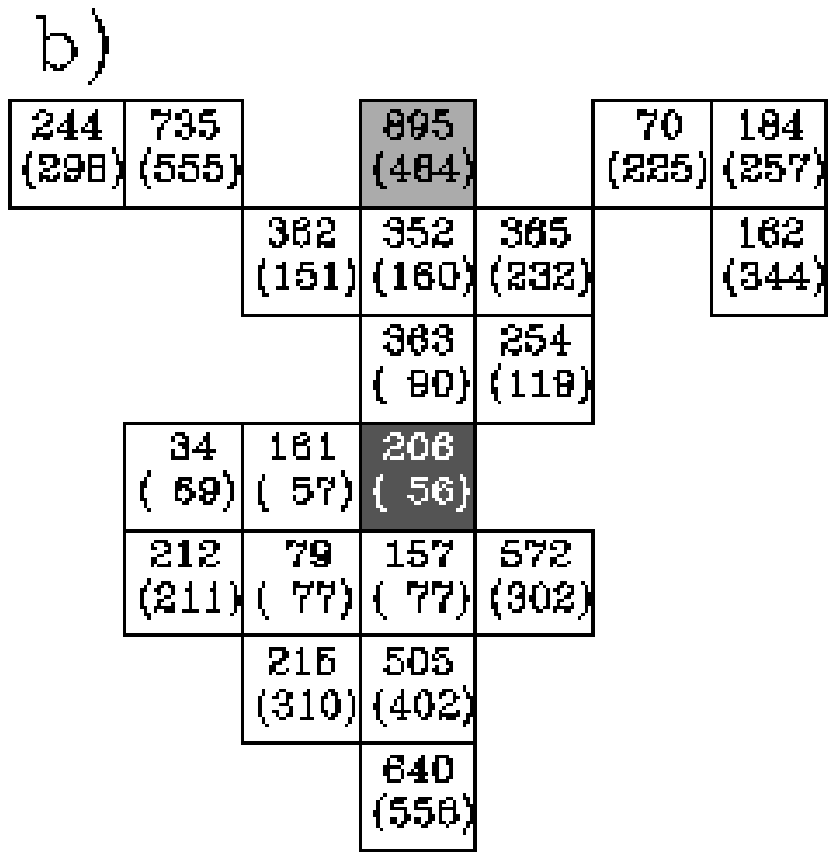,angle=0,width=4.0cm}
\hspace*{1.0cm}\psfig{figure=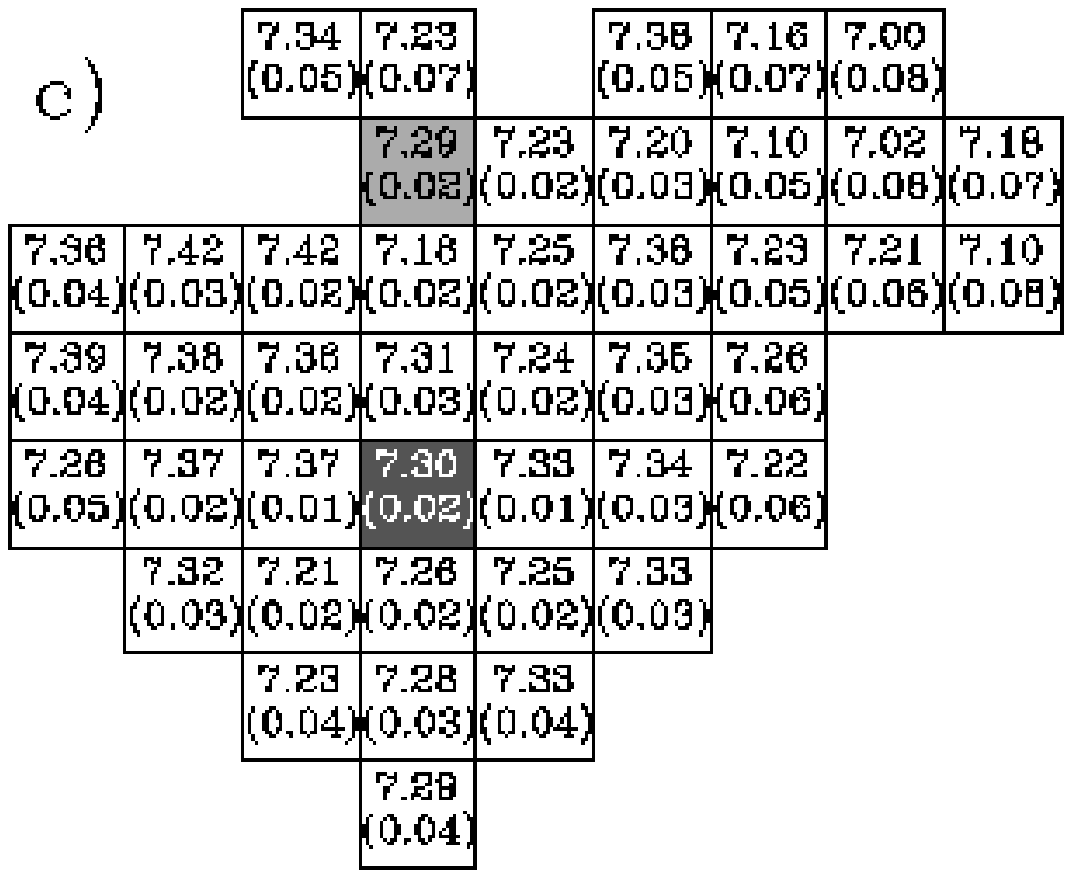,angle=0,width=5.0cm,clip=}
\hspace*{3.0cm}\psfig{figure=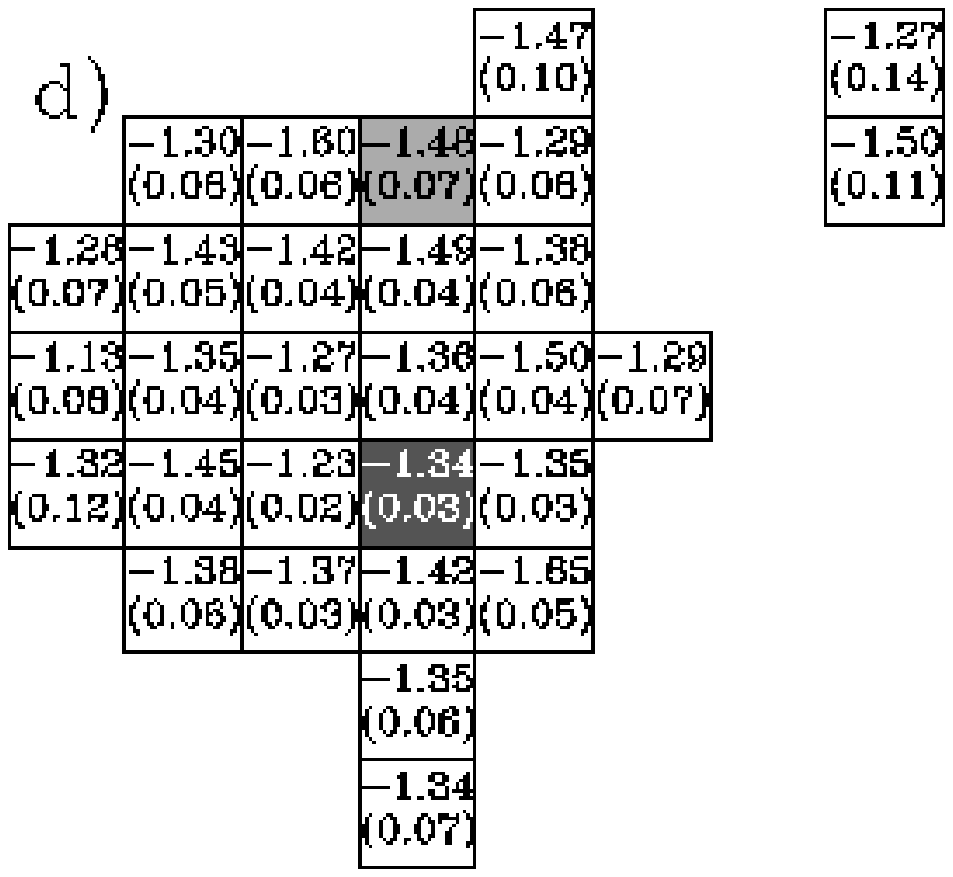,angle=0,width=4.0cm}
\hspace*{1.5cm}\psfig{figure=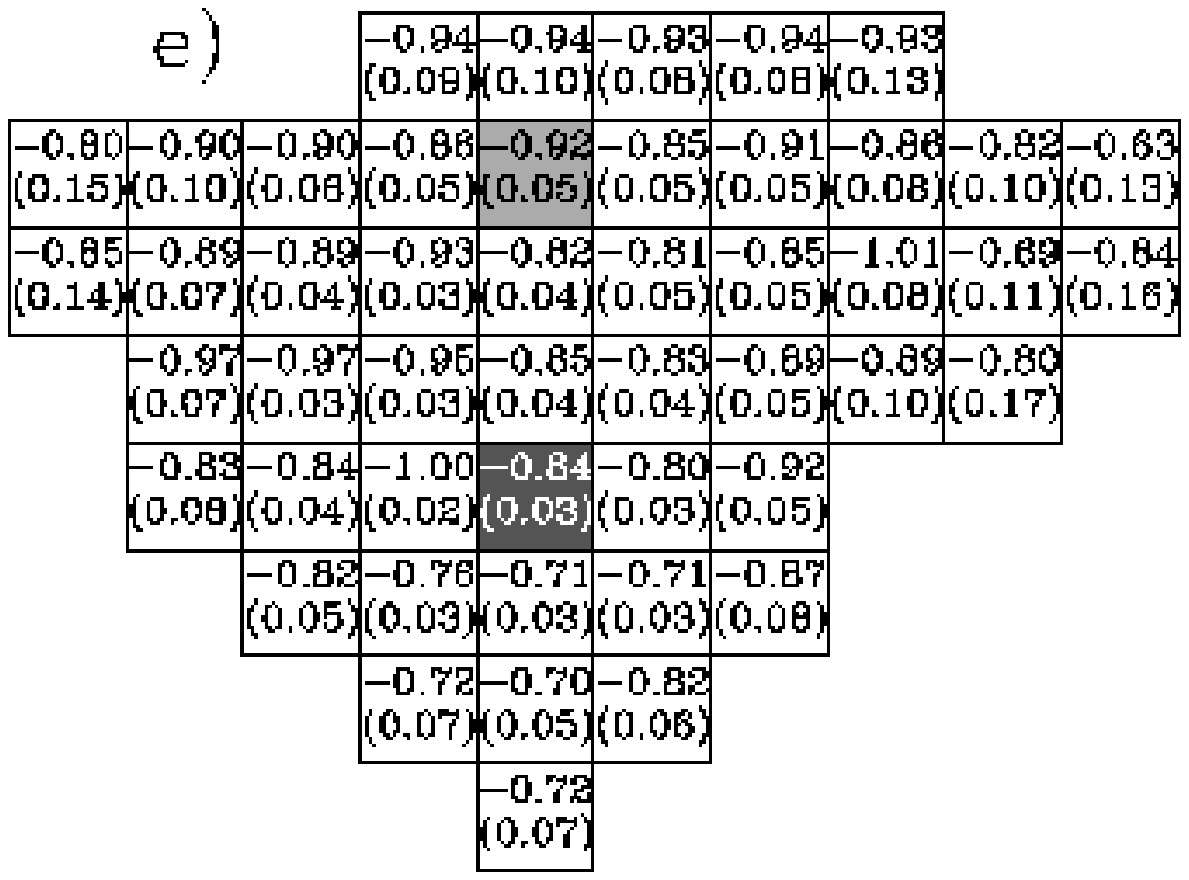,angle=0,width=5.0cm,clip=}
\caption{(a) Electron temperature distribution (in 10$^{4}$K) from the 
[O {\sc iii}] $\lambda$4363/($\lambda$4959+$\lambda$5007) line ratio. 
(b) Electron number density distribution (in cm$^{-3}$) from the 
[S {\sc ii}] $\lambda$6717/$\lambda$6731 line ratio.
(c) Oxygen abundance 12+log O/H distribution. (d) Distribution of
the nitrogen-to-oxygen abundance ratio. (e) Distribution of the neon-to-oxygen
abundance ratio. The dark region is
the region with maximum intensity of H$\alpha$ 6563\AA\ emission line
coincident with the location of clusters 1 and 2. The grey region corresponds to 
clusters 4 and 5. 
}
\label{fig10}
\end{figure*}

\subsection{Distribution of the Physical Parameters and Oxygen Abundance}

The VLT/GIRAFFE panoramic spectra allow also to study the distribution
of the electron temperature $T_e$(O {\sc iii}), the electron number density
$N_e$(S {\sc ii}) and heavy element abundances in the H {\sc ii} region. 
For this we use the spectra
obtained for the 0\farcs52$\times$0\farcs52 apertures. We took in 
consideration only spectra in which at least the following lines of
heavy elements are detected: [O {\sc ii}] $\lambda$3726, 3729\AA, 
[O {\sc iii}] $\lambda$4363, 4959, 5007\AA. This allows to derive
the electron temperature $T_e$(O  {\sc iii}) and oxygen abundance. From these
spectra we excluded those spectra, where the oxygen abundance 
12 + log O/H is derived with an error greater than 0.1 dex.

In Fig. \ref{fig10}a is shown the distribution of the electron temperature
$T_e$(O {\sc iii}). It is seen that the H {\sc ii} region is hot in all
small apertures and has the characteristic temperature of $\sim$ 20000K. 
There is a slight spatial trend of the electron temperature with
$T_e$(O {\sc iii}) being
slightly higher in the western part and slightly lower in the eastern part.
The electron number density derived from the 
[S {\sc ii}] $\lambda$6717, 6731\AA\ emission lines is high, 
of several hundred particles per cm$^3$ (Fig. \ref{fig10}b). However, the
errors in the determination of $N_e$ are large and are 
caused by the low intensity of the [S {\sc ii}] 
emission lines. Similar number densities are derived from the
[O {\sc ii}] $\lambda$3726/$\lambda$3729 flux ratio. Although 
[O {\sc ii}] $\lambda$3726, 3729\AA\ emission lines are brighter than
[S {\sc ii}] $\lambda$6717, 6731\AA, the low signal-to-noise ratio of the
spectrum containing [O {\sc ii}] lines (Fig. \ref{fig2}) due to lower 
sensitivity of the GIRAFFE detector in that wavelength range
prevents the determination
of the electron number density from the [O {\sc ii}] lines in a region
larger than that from the [S {\sc ii}] emission lines. 
As already mentioned, other emission lines detected in our spectra
can in principle be used to determine the electron density.
[Ar {\sc iv}] $\lambda$4711 and 4740\AA\
are strong enough only in the brightest region of SBS 0335--052E, where they
indicate consistently an electron number density in the range 
10$^3$ -- 10$^4$ cm$^{-3}$.
[Cl {\sc iii}] $\lambda$5517 and 5537\AA\ are too weak even
in the brightest region (Fig. \ref{fig2}) and are therefore not used.

The oxygen abundance 12 + log O/H distribution is shown in Fig. \ref{fig10}c.
There are some variations of the oxygen abundance in the range 7.00 -- 7.42
with a slight trend of decreasing of 12 + log O/H from the East to the West.
In particular, it appears that the oxygen abundance in cluster 7 of $\sim$7.2
(western side of Fig. \ref{fig10}) is slightly lower than in other clusters,
confirming the finding by \citet{P06}. 
Thus, there is some evidence for a possible self-enrichment by heavy elements 
\citep[cf.][]{I97b,I99} or for the presence of ``initial'' abundance
variations in the gas. However, we point out here that the errors in
the electron temperature, electron number density and oxygen abundance include
only errors derived based on the photon statistics of non-flux-calibrated 
emission line fluxes and they do not take into account uncertainties in 
pointing, variable seeing, differential refraction, etc., which is difficult to
estimate. Therefore, variations in the oxygen abundance may not be 
statistically significant.

In Fig. \ref{fig10}d are shown variations of N/O abundance ratio. The log N/O
varies in the wide enough range from --1.13 to --1.65 with relatively small
errors. However, the real errors might be much higher because of the 
limitations introduced by small apertures of 0\farcs52$\times$0\farcs52 for
each spectrum. The same is true for the distribution of the Ne/O abundance 
ratio (Fig. \ref{fig10}e).

\subsection{Helium Abundance from the Integrated Spectrum}

SBS 0335--052E being one of the most metal-deficient BCDs plays an important
role in the determination of the primordial He mass fraction $Y_p$ and, thus, 
in the determination of the baryonic mass fraction
of the Universe. Since the precision in the
determination of $Y_p$ should be better than $\sim$1\% to put useful
constraints on the cosmological models, high signal-to-noise spectra
are needed for this. Additionally, several systematic effects should
be taken into account, and spectra and emission line fluxes should be
corrected for them \citep[see for details][]{IT04,ITS06}.
These are the corrections for (1) interstellar extinction, (2)  ionisation
structure, (3)  collisional excitation of helium lines, (4)
fluorescence in helium lines, (5) temperature variations, 
(6)  underlying stellar
absorption, (7)  collisional excitation of hydrogen lines. All
these corrections are at a level of a few percent and, apart from (2)
influence each other in a complicated way. 
The case of SBS 0335--052E is particularly complicated, because
its H {\sc ii} region is dense, hot and optically thick in some He {\sc i}
emission lines \citep{I99}. Therefore, effects (3), (4) and (7) are strong
in the H {\sc ii} region of SBS 0335--052E.

To derive the He abundance we use the integrated spectrum of SBS 0335--052E
because it is least dependent on the observational parameters discussed 
above. The He$^+$ abundance $y^+$ which is derived from the He {\sc i}
emission lines depends on the adopted He {\sc i}
emissivities. We adopt the new He {\sc i} emissivities by \citet{P05}. 
In this paper, following \citet{ITL94,ITL97}, \citet{IT98,IT04} and
\citet{ITS06} we use the five 
strongest He {\sc i} $\lambda$3889, $\lambda$4471, 
$\lambda$5876, $\lambda$6678 and $\lambda$7065 emission lines to derive 
$N_e$(He$^+$) and $\tau$($\lambda$3889).
The He {\sc i} $\lambda$3889 and 
$\lambda$7065 lines play an important role because they are particularly 
sensitive to both quantities. Since the
He {\sc i} $\lambda$3889 line is blended with the H8 $\lambda$3889 line, 
we have subtracted the latter, assuming its intensity to be equal to 0.107 
$I$(H$\beta$) \citep{A84}.

Besides the emissivities the derived $y^+$$\equiv$He$^+$/H$^+$ abundance 
depends on several 
other parameters: collisional excitation of hydrogen emission lines, electron
number density $N_e$(He$^+$) and electron temperature $T_e$(He$^+$), equivalent
widths EW($\lambda$3889), EW($\lambda$4471), EW($\lambda$5876), 
EW($\lambda$6678) and EW($\lambda$7065) of He {\sc i} stellar absorption lines,
optical depth 
$\tau$($\lambda$3889) of the He {\sc i} $\lambda$3889 emission line. 
We use Monte Carlo simulations for the $y^+$ determination randomly varying the parameters
in their ranges. First, we subtract the fractions of the H$\alpha$ and H$\beta$
observed fluxes due to the collisional excitation. We adopt that 
the fraction $\Delta$$I_{coll}$(H$\alpha$)/$I$(H$\alpha$) of the 
H$\alpha$ flux due to the collisional excitation varies in the range 
0\% -- 5\% of the total flux. This fraction is 
randomly generated 100 times in the adopted range. The fraction of
the H$\beta$ emission line flux due to the collisional excitation
is adopted to be three times less than that of the H$\alpha$ flux. 
For each generated fraction
of the H$\alpha$ and H$\beta$ they are subtracted from the total observed
fluxes and then all emission line fluxes are corrected for the interstellar
extinction and abundances of elements are calculated.

To calculate $y^+$ we simultaneously and randomly vary $N_e$(He$^+$),
$T_e$(He$^+$) and $\tau$($\lambda$3889). The
total number of such realizations is 10$^5$ for
each value of $\Delta$$I_{coll}$(H$\alpha$)/$I$(H$\alpha$).
Thus, the total number of Monte Carlo realizations is 
100 $\times$ 10$^5$ = 10$^7$.
As for He {\sc i} underlying stellar absorption, we keep constant values of
EW($\lambda$3889), EW($\lambda$4471), EW($\lambda$5876), EW($\lambda$6678) and 
EW($\lambda$7065) during simulations.
In our spectrum other He {\sc i} emission lines, namely
He {\sc i} $\lambda$3820, $\lambda$4388, $\lambda$4026, $\lambda$4921,
and $\lambda$7281 are seen. However, we do not attempt to use these lines for He 
abundance determination because they are much weaker as compared to the five 
brightest lines, and hence have larger uncertainties. 

We solve the problem by minimization of the expression
\begin{equation}
\chi^2=\sum_i^n\frac{(y^+_i-y^+_{\rm mean})^2}{\sigma^2(y^+_i)}\label{eq1},
\end{equation}
where $y^+_i$ is the He$^+$ abundance derived from the flux of the He {\sc i}
emission line with label $i$, $\sigma(y^+_i)$ is the statistical error
of the He abundance. The quantity $y^+_{\rm mean}$ is the weighted 
mean of the He$^+$
abundance as derived from the equation
\begin{equation}
y^+_{\rm mean}=\frac{\sum_i^k{y^+_i/\sigma^2(y^+_i)}}
{\sum_i^k{1/\sigma^2(y^+_i)}}\label{eq2}.
\end{equation}
We use all five He {\sc i} emission lines to calculate $\chi^2$ (i.e.,
$n$ = 5), but only three lines, He {\sc i} $\lambda$4471,
He {\sc i} $\lambda$5876 and He {\sc i} $\lambda$6678 to calculate
$y^+_{\rm mean}$ ($k$ = 3). This is because the fluxes of 
He {\sc i} $\lambda$3889 and He {\sc i} $\lambda$7065 emission lines are
more uncertain compared to other three He {\sc i} emission lines.

The best solution for $y^+_{\rm mean}$ is found from the minimum of $\chi^2$,
the systematic error $\sigma_{\rm sys}$ is obtained from the dispersion
of $y^+_{\rm mean}$ in the range of $\chi^2$ between $\chi^2_{\rm min}$ and
$\chi^2_{\rm min}$ + 1. Then the total error for $y^+_{\rm mean}$ is
derived from $\sigma^2_{\rm tot}$ = $\sigma^2_{\rm stat}$ + 
$\sigma^2_{\rm sys}$.

   Additionally, since the nebular He {\sc ii} $\lambda$4686\AA\  
emission line was detected, we have added the abundance of doubly ionised 
helium $y^{2+}$ $\equiv$ He$^{2+}$/H$^+$ to $y^+$. Although the He$^{2+}$ zone
is hotter than the He$^{+}$ zone, we adopt $T_e$(He$^{2+}$) = $T_e$(He$^{+}$).
The last assumption introduces a little change in $y$ value, because
the value of $y$$^{2+}$ is small ($\sim$ 4\% of $y^+$).
Finally, the ionisation correction factor $ICF$(He) is taken into account
from \citet{ITS06} to convert $y^+$+$y^{2+}$ to the total He abundance
$y$$\equiv$He/H.

The derived parameters and He abundances for every He {\sc i} emission
line are shown in Table \ref{tab5}. Here,
$\Delta$$I_{coll}$(H$\alpha$)/$I$(H$\alpha$) is the fraction of H$\alpha$
flux due to the collisional excitation. It is seen from this Table, that the
electron number density $N_e$ in the He$^+$ zone is high, 295 cm$^{-3}$, and
is consistent with the number density derived from the [S {\sc ii}] emission
lines. The electron temperature $T_e$(O {\sc iii}) in Table \ref{tab5}
slightly differs from that in Table \ref{tab3} (right column). This is because
in Table \ref{tab5} collisional excitation of hydrogen lines is taken into
account resulting in smaller $C$(H$\beta$) as compared to that in Table
\ref{tab3}. The electron temperature $T_e$(He$^+$) is only slightly lower than
$T_e$(O {\sc iii}) suggesting that temperature fluctuations in the H {\sc ii}
region of SBS 0335--052E are small.
Note, that the optical depth $\tau$($\lambda$3889) of 2.9 is high compared to
that in other BCDs \citep[see e.g. ][]{IT04} implying important contribution 
of the fluorescent enhancement of He {\sc i} emission lines.

The derived weighted mean He mass fraction in SBS 0335--052E, 
$Y$ = 0.2463 $\pm$ 0.0030, is slightly lower (but consistent within the errors) 
than the primordial He mass 
fraction $Y_p$ = 0.24815 $\pm$ 0.0003 $\pm$ 0.0006(syst.) 
from the 3-year data of the WMAP experiment
\citep{S06}
and the D abundance, supporting the standard cosmological model of the primordial nucleosynthesis.

\section{Wolf-Rayet Stars \label{sec:wr}}


\begin{table}[t]
\caption{Helium Abundance \label{tab5}}
\begin{tabular}{lc} \hline
Parameter                           & Value \\ \hline
$T_e$(O {\sc iii}), K               &20180   \\
$T_e$(He$^+$), K                    &20010   \\
$N_e$(He$^+$), cm$^{-3}$            &295     \\
EW$_{abs}$($\lambda$4471), \AA      &0.4     \\
EW$_{abs}$($\lambda$3889)/EW$_{abs}$($\lambda$4471)      &1.0     \\
EW$_{abs}$($\lambda$5876)/EW$_{abs}$($\lambda$4471)      &0.3     \\
EW$_{abs}$($\lambda$6678)/EW$_{abs}$($\lambda$4471)      &0.1     \\
EW$_{abs}$($\lambda$7065)/EW$_{abs}$($\lambda$4471)      &0.1     \\
$\tau$($\lambda$3889)   &2.9 \\
$\Delta$$I_{coll}$(H$\alpha$)/$I$(H$\alpha$), \% &4.95 \\
$\chi^2_{min}$          &0.1825 \\
$y^+$($\lambda$3889)    &0.0782$\pm$0.0134 \\
$y^+$($\lambda$4471)    &0.0802$\pm$0.0032 \\
$y^+$($\lambda$5876)    &0.0788$\pm$0.0012 \\
$y^+$($\lambda$6678)    &0.0787$\pm$0.0018 \\
$y^+$($\lambda$7065)    &0.0789$\pm$0.0018 \\
$y^+$(weighted mean)    &0.0789$\pm$0.0010 \\
$y^{++}$($\lambda$4686) &0.0034$\pm$0.0002 \\
$ICF$(He)               &0.993             \\
$y$(weighted mean)      &0.0817$\pm$0.0010 \\
$Y$(weighted mean)      &0.2463$\pm$0.0030 \\
\hline
\end{tabular}
\end{table}

\begin{figure}
\hspace*{0.0cm}\psfig{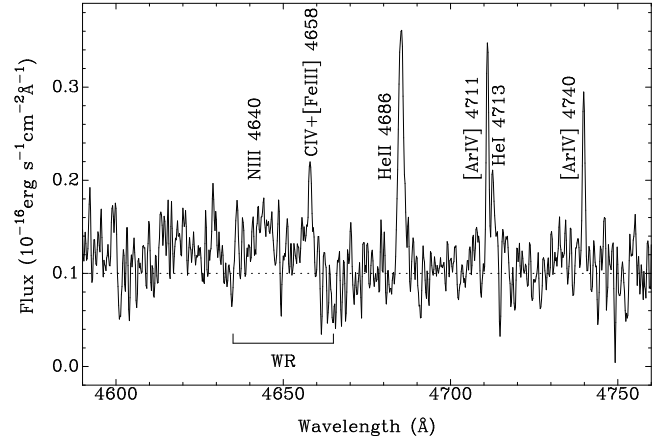}
\caption{Part of the spectrum of cluster 3 showing the probable broad Wolf-Rayet
emission lines N {\sc iii} $\lambda$4640 and C {\sc iv} $\lambda$4658
labelled ``WR''. }
\label{fig11}
\end{figure}

The search for Wolf-Rayet (WR) stars in extremely low-metallicity dwarf
galaxies is of great interest for constraining stellar evolution models.
However, such studies are difficult since the strength of WR emission
lines is significantly reduced with decreasing metallicity. Therefore,
very high signal-to-noise ratio spectra are required to detect weak
WR features. For a long time no WR galaxies with an oxygen abundance
12 + log O/H $<$ 7.9 were known \citep{M91}. Later, \citet{I97c} and
\citet{L97} have discovered WR stars in I Zw 18, at that time the most 
metal-deficient emission-line galaxy with the oxygen abundance 
12 + log O/H = 7.17. Thus, it appears that WR stars could be found in
any other dwarf emission-line galaxy with active star formation,
including SBS 0335--052E. However, the strength of WR emission features
depends not only on metallicity but also on the age of a starburst.
In starbursts with the metallicity of SBS 0335--052E, the WR stage is
expected to be very short, typically less than 1 Myr \citep{SV98,DM98}. 
Therefore, not all young clusters
ionising the interstellar medium in SBS 0335--052E may be expected to contain
WR stars. Recently, \citet{P06} have found that WR stars of the early
carbon sequence (WC4 stars) are present in cluster 3 
of SBS 0335--052E.

GIRAFFE/ARGUS observations allow in principle a more detailed search for
WR stars in SBS 0335--052E, a more precise localisation in the galaxy,
and also to resolve the $\lambda$4650 WR bump into N {\sc iii} $\lambda$4640
and C {\sc iv} $\lambda$4658 broad features, thus allowing the detection
of both late nitrogen WR stars (WNL stars) and early carbon WR stars
(WCE stars). Previous observations by \citet{P06} had too low spectral
resolution to definitely distinguish between two types of WR stars.
However, there are some limitations of GIRAFFE/ARGUS observations
which make such study more difficult
as compared with that of \citet{P06}. Although \citet{P06} have observed
with the smaller 3.6m ESO telescope, their spectrum has a higher S/N ratio
because of the $\sim$ 3 times longer exposure and $\sim$ 10 times lower
spectral resolution. 

We checked all 0\farcs52$\times$0\farcs52 spectra obtained with GIRAFFE
and found that broad WR features near $\lambda$4650 are likely present 
only in one spectrum associated with cluster 3. This spectrum is shown
in Fig. \ref{fig11}. The S/N ratio of $\sim$ 5 of this 
spectrum is not high, but broad WR features are clearly seen.
Thus, we confirm the finding by \citet{P06} that
WR stars appear to be present in cluster 3. However, we find that
the WR feature consists in fact of two lines: N {\sc iii} $\lambda$4640 and
C {\sc iv} $\lambda$4658. The latter line is blended with the much
narrower nebular [Fe {\sc iii}] $\lambda$4658 emission line. 
We find that, after subtraction of the [Fe {\sc iii}] line from the $\lambda$4658
blend, the fluxes of the N {\sc iii} $\lambda$4640 and C {\sc iv} $\lambda$4658
lines and their FWHMs are similar, 
$\sim$ (4$\pm$1)$\times$10$^{-17}$ erg s$^{-1}$ cm$^{-2}$ 
and 6.5\AA, respectively.
Thus, the total flux of the $\lambda$4650 bump ( N {\sc iii} $\lambda$4640 +
C {\sc iv} $\lambda$4658) non-corrected for extinction
is $\sim$8$\times$10$^{-17}$ erg s$^{-1}$ cm$^{-2}$, or 2/3 that measured
by \citet{P06} in a larger aperture. The total equivalent width of this
bump is EW($\lambda$4650) $\sim$ 9\AA. The 
C {\sc iv} $\lambda$4658 emission line should be accompanied by
the C {\sc iv} $\lambda$5808 emission line with a comparable flux.
Unfortunately, the redshifted C {\sc iv} $\lambda$5808 emission line 
in SBS 0335--052E coincides with the night sky Na {\sc i} 
$\lambda$5890, 5895 emission lines. Therefore, the imperfect night sky 
subtraction hinders the detection of the WR line.

The observed (i.e.\ not extinction corrected)
N {\sc iii} $\lambda$4640 and C {\sc iv} $\lambda$4658 
emission line luminosities of cluster 3 are
equal to $L$(N {\sc iii} $\lambda$4640) = $L$(C {\sc iv} $\lambda$4658) = 
1.4$\times$10$^{37}$ erg s$^{-1}$ and correspond to the number of WNL and
stars $N$(WNL) $\sim$ 35$\pm$8 and to the number of WC4 stars 
$N$(WC4) $\sim$ 3$\pm$1 
if we adopt the ``standard'' WR line luminosities computed by  \citet{SV98},
i.e.\ assuming N {\sc iii} $\lambda$4640 and C {\sc iv} $\lambda$4658 emission line 
luminosities of 4$\times$10$^{35}$ erg s$^{-1}$ 
and 5$\times$10$^{36}$ erg s$^{-1}$ respectively for a single WNL and WC4 star.
The derived values of WNL and WC4 stars are to be taken with a grain of salt
for various reasons. First these are likely
lower limits because of the neglected reddening, and since WR stars at low metallicity
may have lower instrinsic line luminosities \citep[e.g. ][]{CH06}.
Second, it is not necessarily clear that the observed lines are indeed due to 
WN and WC stars, as their strengths and widths are somewhat unusual.
Their relatively small widths could be due to lower mass loss rates
and/or smaller wind velocities in WR stars at such low metallicity
\citep{C02,VK05,CH06}.
Alternatively, from their relative strength and the small widths,
the observed lines could also correspond to very late WNL stars 
(WN 10h-11h) \citep[cf.][]{CS97}, if most of the C {\sc iv} $\lambda$4658 
was nebular.
Given the faintness of these spectral signatures and the lack of known individual WR stars 
at such low metallicities as comparison objects, it is difficult to draw
firm conclusions on the WR content in this cluster.

\section{Conclusions \label{sec:concl}}

In this paper we present panoramic spectroscopic observations with
the VLT/GIRAFFE in the spectral range $\lambda$3620--9400\AA\ of one of the 
most metal-deficient blue compact dwarf (BCD)
galaxies, SBS 0335--052E. Our findings can be summarized as follows:

1. The morphology of the galaxy in different lines is very similar, except for the He {\sc ii} 
$\lambda$4686\AA\ line, suggesting that the main ionising source
in this galaxy is the compact super-star clusters (SSC) 1+2 in the SE part of 
the galaxy, most likely the cluster 1. 
The equivalent number of O7{\sc v} ionising stars in this cluster
is $\sim$13000, making it the most powerful SSC known so far.

2. The maximum emission of the He {\sc ii} $\lambda$4686\AA\ line is offset to
the older compact clusters 4 and 5. Additionally, its width is 1.5 -- 2
times greater than the width of other lines. Furthermore, the width of this and
other emission lines is much higher in clusters 4,5 than in cluster 1.
These facts imply that the hard ionising radiation responsible for He {\sc ii}
emission is most likely produced by fast radiative shocks.
However, some contribution from WR stars located in cluster 3 cannot be
excluded.

3. The analysis of kinematical properties of the ionised gas in SBS 0335--052E
suggests the presence of gas outflow in the direction perpendicular to the
galaxy disk.

4. We derive physical parameters in the H {\sc ii} region of SBS 0335--052E,
suggesting that the ionised gas is hot ($\sim$20000K) and dense ($\ga$ several
hundred particles per cm$^{-3}$). The oxygen abundances 12 + log O/H in the
brightest region and in the integrated spectrum of SBS 0335--052E are
respectively 7.31$\pm$0.01 and 7.29$\pm$0.02, and they are
consistent with previous
determinations. The other heavy element-to-oxygen abundance ratios are
consistent with the average values derived for other most metal-deficient
galaxies. There is a slight decrease of oxygen abundance from the East to the
West suggesting some self-enrichment of the ionised gas by heavy 
elements.

5. We derive the He mass fraction $Y$ from the integrated spectrum taking into
account all possible systematic effects in the He abundance determination.
Our value $Y$ = 0.2463 $\pm$ 0.0030 is slightly lower but consistent within
the errors with the primordial He mass 
fraction determined from the 3 year WMAP data and from primordial deuterium measurements.

6. We confirm the presence of Wolf-Rayet stars in the cluster 3 found 
previously by \citet{P06}. The lower limits of the WNL and WC4 star numbers
are $\sim$ 35$\pm$8 and $\sim$ 3$\pm$1, respectively
if standard WR line luminosities are assumed.

\begin{acknowledgements}
Y. I. I. and N. G. G. thank the hospitality of the Observatoire de Gen\`eve.
DS thanks Paul Crowther for useful comments on WR stars and the 
Swiss National Science Foundation for support.
\end{acknowledgements}

\Online


{
\begin{longtable}{lrrr} 
\caption{\label{tab2}Emission Line Fluxes and Equivalent Widths in the 
Spectrum of the Brightest Region}\\
\hline
\endfirsthead
\caption{Continued.}\\
\hline Line            &
100$\times$$F$($\lambda$)/$F$(H$\beta$)&
100$\times$$I$($\lambda$)/$I$(H$\beta$)&
$EW$($\lambda$)$^{\rm a}$ \\
\hline
\endhead
\hline
Line            &
100$\times$$F$($\lambda$)/$F$(H$\beta$)&
100$\times$$I$($\lambda$)/$I$(H$\beta$)&
$EW$($\lambda$)$^{\rm a}$ \\ \hline
\multicolumn{4}{c}{LR1 ($\lambda$3620--4081)} \\ \hline
3634 He {\sc i}        &  0.25$\pm$0.02&  0.29$\pm$0.13&   0.33$\pm$0.04 \\
3631 H24        &  0.47$\pm$0.02&  1.08$\pm$0.23&   0.89$\pm$0.06 \\
3674 H23        &  0.59$\pm$0.03&  1.21$\pm$0.20&   1.13$\pm$0.07 \\
3676 H22        &  0.60$\pm$0.03&  1.21$\pm$0.19&   1.19$\pm$0.07 \\
3679 H21        &  0.80$\pm$0.04&  1.36$\pm$0.15&   1.86$\pm$0.08 \\
3683 H20        &  1.16$\pm$0.04&  1.76$\pm$0.13&   2.85$\pm$0.10 \\
3687 H19        &  0.83$\pm$0.04&  1.33$\pm$0.13&   2.28$\pm$0.10 \\
3692 H18        &  1.08$\pm$0.04&  1.62$\pm$0.12&   2.90$\pm$0.09 \\
3697 H17        &  1.07$\pm$0.04&  1.70$\pm$0.14&   2.31$\pm$0.10 \\
3704 H16        &  1.40$\pm$0.04&  2.06$\pm$0.12&   3.17$\pm$0.09 \\
3705 He {\sc i}        &  0.59$\pm$0.03&  0.68$\pm$0.08&   1.37$\pm$0.08 \\
3712 H15        &  1.21$\pm$0.04&  1.83$\pm$0.13&   2.81$\pm$0.09 \\
3722 H14        &  1.80$\pm$0.05&  2.53$\pm$0.13&   3.90$\pm$0.10 \\
3726 [O {\sc ii}]      &  8.12$\pm$0.13&  9.26$\pm$0.18&  17.73$\pm$0.14 \\
3729 [O {\sc ii}]      &  9.51$\pm$0.15& 10.85$\pm$0.20&  20.86$\pm$0.17 \\
3734 H13        &  2.01$\pm$0.05&  2.76$\pm$0.13&   4.46$\pm$0.13 \\
3750 H12        &  2.75$\pm$0.07&  3.59$\pm$0.14&   6.12$\pm$0.15 \\
3771 H11        &  3.08$\pm$0.08&  3.96$\pm$0.15&   6.65$\pm$0.14 \\
3798 H10        &  4.03$\pm$0.10&  5.04$\pm$0.17&   8.45$\pm$0.19 \\
3820 He {\sc i}        &  0.72$\pm$0.07&  0.81$\pm$0.11&   1.63$\pm$0.17 \\
3835 H9         &  6.19$\pm$0.15&  7.37$\pm$0.20&  15.35$\pm$0.29 \\
3868 [Ne {\sc iii}]    & 21.29$\pm$0.36& 23.84$\pm$0.43&  47.27$\pm$0.45 \\
3889 He {\sc i}+H8     & 14.13$\pm$0.29& 16.22$\pm$0.36&  32.49$\pm$0.50 \\
3965 He {\sc i}        &  0.61$\pm$0.23&  0.68$\pm$0.26&   1.60$\pm$0.58 \\
3967 [Ne {\sc iii}]    &  6.51$\pm$0.29&  7.20$\pm$0.33&  18.31$\pm$0.70 \\
3970 H7         & 14.02$\pm$0.42& 15.85$\pm$0.50&  39.08$\pm$0.93 \\ \hline
\multicolumn{4}{c}{LR2 ($\lambda$3964--4567)} \\ \hline
3965 He {\sc i}        &  0.64$\pm$0.01&  0.71$\pm$0.08&   1.30$\pm$0.04 \\
3967 [Ne {\sc iii}]    &  7.16$\pm$0.11&  7.91$\pm$0.15&  14.64$\pm$0.08 \\
3970 H7         & 13.97$\pm$0.20& 15.92$\pm$0.26&  28.70$\pm$0.10 \\
4026 He {\sc i}        &  1.60$\pm$0.03&  1.75$\pm$0.09&   3.38$\pm$0.05 \\
4068 [S {\sc ii}]      &  0.25$\pm$0.02&  0.28$\pm$0.08&   0.54$\pm$0.04 \\
4101 H$\delta$         & 23.99$\pm$0.35& 26.53$\pm$0.42&  51.74$\pm$0.16 \\
4121 He {\sc i}        &  0.24$\pm$0.02&  0.26$\pm$0.08&   0.52$\pm$0.05 \\
4144 He {\sc i}        &  0.41$\pm$0.03&  0.44$\pm$0.08&   0.89$\pm$0.06 \\
4169 He {\sc i}        &  0.10$\pm$0.02&  0.10$\pm$0.08&   0.21$\pm$0.05 \\
4227 [Fe {\sc v}]      &  0.20$\pm$0.03&  0.21$\pm$0.08&   0.45$\pm$0.08 \\
4249 [Fe {\sc ii}]     &  0.07$\pm$0.02&  0.08$\pm$0.07&   0.17$\pm$0.05 \\
4287 [Fe {\sc ii}]     &  0.16$\pm$0.02&  0.17$\pm$0.07&   0.39$\pm$0.07 \\
4340 H$\gamma$         & 46.80$\pm$0.68& 49.82$\pm$0.75& 114.10$\pm$0.33 \\
4359 [Fe {\sc ii}]     &  0.17$\pm$0.03&  0.18$\pm$0.06&   0.59$\pm$0.07 \\
4363 [O {\sc iii}]     & 10.52$\pm$0.17& 11.08$\pm$0.19&  25.39$\pm$0.18 \\
4368 O {\sc i}         &  0.10$\pm$0.03&  0.11$\pm$0.07&   0.25$\pm$0.06 \\
4379 N {\sc iii}       &  0.10$\pm$0.03&  0.10$\pm$0.07&   0.24$\pm$0.08 \\
4388 He {\sc i}        &  0.44$\pm$0.03&  0.46$\pm$0.07&   1.10$\pm$0.10 \\
4414 [Fe {\sc ii}]     &  0.08$\pm$0.02&  0.08$\pm$0.07&   0.19$\pm$0.07 \\
4416 [Fe {\sc ii}]     &  0.05$\pm$0.02&  0.05$\pm$0.06&   0.13$\pm$0.07 \\
4438 He {\sc i}        &  0.11$\pm$0.03&  0.12$\pm$0.07&   0.29$\pm$0.08 \\
4471 He {\sc i}        &  3.75$\pm$0.08&  3.90$\pm$0.10&   9.83$\pm$0.15 \\ \hline
\multicolumn{4}{c}{LR3 ($\lambda$4501--5078)} \\ \hline
4452 [Fe {\sc ii}]     &  0.10$\pm$0.01&  0.10$\pm$0.06&   0.25$\pm$0.04 \\
4471 He {\sc i}        &  3.75$\pm$0.06&  3.90$\pm$0.09&   9.88$\pm$0.08 \\
4541 He {\sc ii}       &  0.05$\pm$0.01&  0.05$\pm$0.06&   0.13$\pm$0.04 \\
4571 Mg {\sc i}        &  0.09$\pm$0.01&  0.10$\pm$0.06&   0.26$\pm$0.04 \\
4658 [Fe {\sc iii}]    &  0.26$\pm$0.02&  0.26$\pm$0.06&   0.75$\pm$0.05 \\
4686 He {\sc ii}       &  1.39$\pm$0.03&  1.41$\pm$0.06&   4.36$\pm$0.08 \\
4702 [Fe {\sc iii}]    &  0.12$\pm$0.01&  0.12$\pm$0.05&   0.39$\pm$0.05 \\
4711 [Ar {\sc iv}]     &  1.08$\pm$0.02&  1.09$\pm$0.06&   3.34$\pm$0.06 \\
4713 He {\sc i}        &  0.70$\pm$0.02&  0.70$\pm$0.05&   2.14$\pm$0.06 \\
4734 [Fe {\sc iii}]    &  0.04$\pm$0.01&  0.04$\pm$0.05&   0.13$\pm$0.04 \\
4740 [Ar {\sc iv}]     &  0.95$\pm$0.03&  0.96$\pm$0.06&   3.04$\pm$0.08 \\
4861 H$\beta$          &100.00$\pm$1.43&100.00$\pm$1.44& 321.70$\pm$0.46 \\
4881 [Fe {\sc iii}]    &  0.10$\pm$0.02&  0.10$\pm$0.05&   0.31$\pm$0.06 \\
4893 [Fe {\sc vii}]    &  0.04$\pm$0.01&  0.04$\pm$0.05&   0.14$\pm$0.05 \\
4907 [Fe {\sc iv}]     &  0.11$\pm$0.02&  0.11$\pm$0.05&   0.41$\pm$0.05 \\
4921 He {\sc i}        &  1.04$\pm$0.03&  1.03$\pm$0.05&   4.05$\pm$0.11 \\
4930 [Fe {\sc iii}]    &  0.09$\pm$0.02&  0.09$\pm$0.04&   0.34$\pm$0.06 \\
4959 [O {\sc iii}]     &107.82$\pm$1.54&106.45$\pm$1.53& 406.70$\pm$0.53 \\
4972 [Fe {\sc vi}]     &  0.11$\pm$0.02&  0.10$\pm$0.04&   0.41$\pm$0.07 \\
4986 [Fe {\sc iii}]    &  0.41$\pm$0.03&  0.40$\pm$0.05&   1.48$\pm$0.08 \\
5007 [O {\sc iii}]     &333.72$\pm$4.75&327.89$\pm$4.70&1075.00$\pm$0.75 \\ \hline
\multicolumn{4}{c}{LR4 ($\lambda$5010--5831)} \\ \hline
4959 [O {\sc iii}]     &107.84$\pm$1.54&106.46$\pm$1.53& 317.50$\pm$0.32 \\
4986 [Fe {\sc iii}]    &  0.41$\pm$0.02&  0.41$\pm$0.05&   1.37$\pm$0.06 \\
5007 [O {\sc iii}]     &325.05$\pm$4.62&319.37$\pm$4.57& 912.80$\pm$0.57 \\
5016 He {\sc i}        &  2.01$\pm$0.04&  1.97$\pm$0.06&   5.99$\pm$0.08 \\
5041 Si {\sc ii}       &  0.15$\pm$0.02&  0.14$\pm$0.05&   0.51$\pm$0.05 \\
5048 He {\sc i}        &  0.21$\pm$0.02&  0.21$\pm$0.05&   0.74$\pm$0.07 \\
5147 [Fe {\sc vi}]     &  0.06$\pm$0.03&  0.05$\pm$0.05&   0.19$\pm$0.06 \\
5158 [Fe {\sc vii}]    &  0.05$\pm$0.02&  0.05$\pm$0.05&   0.18$\pm$0.06 \\
5176 [Fe {\sc vi}]     &  0.11$\pm$0.02&  0.11$\pm$0.05&   0.42$\pm$0.08 \\
5198 [N {\sc i}]       &  0.21$\pm$0.02&  0.20$\pm$0.05&   0.74$\pm$0.09 \\
5200 [N {\sc i}]       &  0.09$\pm$0.02&  0.08$\pm$0.04&   0.31$\pm$0.08 \\
5261 [Fe {\sc ii}]     &  0.06$\pm$0.02&  0.06$\pm$0.04&   0.21$\pm$0.06 \\
5270 [Fe {\sc iii}]    &  0.18$\pm$0.03&  0.17$\pm$0.05&   0.64$\pm$0.10 \\
5273 [Fe {\sc ii}]     &  0.02$\pm$0.01&  0.02$\pm$0.04&   0.09$\pm$0.06 \\
5323 [Cl {\sc iv}]     &  0.05$\pm$0.03&  0.04$\pm$0.05&   0.17$\pm$0.09 \\
5335 [Fe {\sc vi}]     &  0.03$\pm$0.02&  0.03$\pm$0.04&   0.12$\pm$0.09 \\
5411 He {\sc ii}       &  0.14$\pm$0.03&  0.13$\pm$0.04&   0.52$\pm$0.11 \\
5517 [Cl {\sc iii}]    &  0.11$\pm$0.03&  0.10$\pm$0.05&   0.43$\pm$0.11 \\
5537 [Cl {\sc iii}]    &  0.08$\pm$0.02&  0.07$\pm$0.04&   0.31$\pm$0.11 \\
5631 [Fe {\sc vi}]     &  0.02$\pm$0.02&  0.02$\pm$0.04&   0.08$\pm$0.10 \\
5639 [Fe {\sc vi}]     &  0.03$\pm$0.04&  0.03$\pm$0.05&   0.12$\pm$0.09 \\
5721 [Fe {\sc vii}]    &  0.05$\pm$0.03&  0.04$\pm$0.04&   0.21$\pm$0.13 \\ \hline
\multicolumn{4}{c}{LR5 ($\lambda$5741--6524)} \\ \hline
5755 [N {\sc ii}]      &  0.05$\pm$0.01&  0.05$\pm$0.03&   0.24$\pm$0.03 \\
5876 He {\sc i}        & 10.18$\pm$0.15&  9.25$\pm$0.14&  47.95$\pm$0.13 \\
5957 Si {\sc ii}       &  0.04$\pm$0.01&  0.03$\pm$0.03&   0.19$\pm$0.03 \\
5979 Si {\sc ii}       &  0.06$\pm$0.01&  0.05$\pm$0.03&   0.29$\pm$0.03 \\
6046 O {\sc i}         &  0.05$\pm$0.01&  0.05$\pm$0.03&   0.27$\pm$0.04 \\
6102 He {\sc ii}       &  0.04$\pm$0.01&  0.03$\pm$0.03&   0.21$\pm$0.04 \\
6118 He {\sc ii}       &  0.01$\pm$0.01&  0.01$\pm$0.02&   0.08$\pm$0.03 \\
6133 [Fe {\sc iii}]    &  0.02$\pm$0.01&  0.02$\pm$0.02&   0.11$\pm$0.03 \\
6300 [O {\sc i}]       &  0.74$\pm$0.02&  0.65$\pm$0.03&   4.10$\pm$0.06 \\
6312 [S {\sc iii}]     &  0.64$\pm$0.02&  0.56$\pm$0.03&   3.67$\pm$0.07 \\
6348 Si {\sc ii}       &  0.08$\pm$0.01&  0.07$\pm$0.02&   0.46$\pm$0.04 \\
6363 [O {\sc i}]       &  0.27$\pm$0.01&  0.24$\pm$0.02&   1.58$\pm$0.05 \\
6371 Si {\sc ii}       &  0.09$\pm$0.01&  0.08$\pm$0.02&   0.55$\pm$0.05 \\ \hline
\multicolumn{4}{c}{LR6 ($\lambda$6438--7184)} \\ \hline
6363 [O {\sc i}]       &  0.28$\pm$0.01&  0.25$\pm$0.03&   1.50$\pm$0.05 \\
6371 Si {\sc ii}       &  0.09$\pm$0.01&  0.08$\pm$0.03&   0.47$\pm$0.05 \\
6548 [N {\sc ii}]      &  0.44$\pm$0.01&  0.38$\pm$0.03&   1.98$\pm$0.05 \\
6563 H$\alpha$         &317.90$\pm$4.52&274.31$\pm$4.26&1427.00$\pm$0.77 \\
6583 [N {\sc ii}]      &  1.33$\pm$0.02&  1.14$\pm$0.04&   6.23$\pm$0.07 \\
6678 He {\sc i}        &  3.07$\pm$0.05&  2.63$\pm$0.05&  17.76$\pm$0.12 \\
6717 [S {\sc ii}]      &  2.27$\pm$0.04&  1.94$\pm$0.04&  13.41$\pm$0.11 \\
6731 [S {\sc ii}]      &  1.77$\pm$0.03&  1.51$\pm$0.04&  10.49$\pm$0.10 \\
6739 [Fe {\sc iii}]    &  0.06$\pm$0.01&  0.05$\pm$0.02&   0.33$\pm$0.05 \\
7002 O {\sc i}         &  0.07$\pm$0.01&  0.06$\pm$0.02&   0.47$\pm$0.05 \\
7065 He {\sc i}        &  5.98$\pm$0.09&  4.99$\pm$0.09&  38.51$\pm$0.19 \\ \hline
\multicolumn{4}{c}{LR7 ($\lambda$7102--8343)} \\ \hline
7065 He {\sc i}        &  5.98$\pm$0.09&  4.99$\pm$0.09&  31.85$\pm$0.13 \\
7136 [Ar {\sc iii}]    &  1.93$\pm$0.03&  1.60$\pm$0.04&  11.48$\pm$0.10 \\
7171 [Ar {\sc iv}]     &  0.09$\pm$0.01&  0.08$\pm$0.02&   0.59$\pm$0.06 \\
7237 [Ar {\sc iv}]     &  0.06$\pm$0.01&  0.05$\pm$0.02&   0.37$\pm$0.06 \\
7254 O {\sc i}         &  0.11$\pm$0.01&  0.09$\pm$0.02&   0.68$\pm$0.07 \\
7263 [Ar {\sc iv}]     &  0.04$\pm$0.01&  0.03$\pm$0.02&   0.25$\pm$0.06 \\
7281 He {\sc i}        &  0.84$\pm$0.02&  0.70$\pm$0.03&   5.38$\pm$0.10 \\
7320 [O {\sc ii}]      &  0.69$\pm$0.02&  0.57$\pm$0.02&   4.47$\pm$0.10 \\
7330 [O {\sc ii}]      &  0.61$\pm$0.02&  0.50$\pm$0.02&   4.06$\pm$0.10 \\
7468 N {\sc i}         &  0.04$\pm$0.01&  0.03$\pm$0.02&   0.26$\pm$0.07 \\
7751 [Ar {\sc iii}]    &  0.57$\pm$0.02&  0.46$\pm$0.02&   4.18$\pm$0.12 \\
7816 He {\sc i}        &  0.13$\pm$0.01&  0.11$\pm$0.02&   0.96$\pm$0.10 \\
8046 [Cl {\sc iv}]     &  0.08$\pm$0.01&  0.07$\pm$0.02&   0.64$\pm$0.11 \\ \hline
\multicolumn{4}{c}{LR8 ($\lambda$8206--9400)} \\ \hline
8223 N {\sc i}         &  0.06$\pm$0.01&  0.05$\pm$0.02&   0.52$\pm$0.08 \\
8281 P32        &  0.09$\pm$0.01&  0.14$\pm$0.05&   0.77$\pm$0.09 \\
8298 P28        &  0.10$\pm$0.01&  0.16$\pm$0.04&   0.90$\pm$0.09 \\
8306 P27        &  0.09$\pm$0.01&  0.15$\pm$0.05&   0.87$\pm$0.09 \\
8314 P26        &  0.16$\pm$0.01&  0.20$\pm$0.03&   1.51$\pm$0.10 \\
8323 P25        &  0.15$\pm$0.01&  0.19$\pm$0.04&   1.38$\pm$0.11 \\
8334 P24        &  0.28$\pm$0.01&  0.30$\pm$0.03&   2.53$\pm$0.10 \\
8346 P23        &  0.16$\pm$0.01&  0.20$\pm$0.03&   1.52$\pm$0.09 \\
8359 P22        &  0.31$\pm$0.01&  0.32$\pm$0.02&   2.89$\pm$0.11 \\
8362 He {\sc i}        &  0.14$\pm$0.01&  0.11$\pm$0.01&   1.29$\pm$0.09 \\
8374 P21        &  0.37$\pm$0.01&  0.35$\pm$0.02&   3.76$\pm$0.13 \\
8392 P20        &  0.39$\pm$0.01&  0.38$\pm$0.02&   3.53$\pm$0.12 \\
8413 P19        &  0.38$\pm$0.01&  0.38$\pm$0.02&   3.40$\pm$0.12 \\
8438 P18        &  0.39$\pm$0.01&  0.38$\pm$0.02&   3.40$\pm$0.11 \\
8446 O {\sc i}         &  0.84$\pm$0.02&  0.65$\pm$0.02&   7.11$\pm$0.14 \\
8467 P17        &  0.58$\pm$0.02&  0.53$\pm$0.02&   5.09$\pm$0.12 \\
8502 P16        &  0.59$\pm$0.01&  0.54$\pm$0.02&   5.01$\pm$0.14 \\
8545 P15        &  0.75$\pm$0.02&  0.67$\pm$0.03&   5.62$\pm$0.11 \\
8598 P14        &  0.88$\pm$0.02&  0.77$\pm$0.03&   6.73$\pm$0.13 \\
8664 P13        &  1.03$\pm$0.02&  0.89$\pm$0.03&   7.07$\pm$0.11 \\
8751 P12        &  1.43$\pm$0.03&  1.18$\pm$0.03&  11.98$\pm$0.15 \\
8863 P11        &  1.72$\pm$0.03&  1.40$\pm$0.04&  12.72$\pm$0.13 \\
9015 P10        &  1.59$\pm$0.03&  1.29$\pm$0.03&  13.61$\pm$0.15 \\
9069 [S {\sc iii}]     &  4.74$\pm$0.07&  3.58$\pm$0.07&  44.31$\pm$0.25 \\
\hline
$C$(H$\beta$)            &\multicolumn{3}{c}{0.190$\pm$0.018} \\
EW$_{abs}$ (\AA)               &\multicolumn{3}{c}{1.3$\pm$0.2} \\
$F$(H$\beta$) (10$^{-14}$ erg s$^{-1}$ cm$^{-2}$)  &\multicolumn{3}{c}{4.3$\pm$0.2} \\
\hline


\end{longtable}
}

 \par\twocolumn




\begin{table*}[t]
\caption{Emission Line Fluxes and Equivalent Widths in the Integrated Spectrum 
of the H {\sc ii} Region \label{tab4}}
\begin{tabular}{lrrrrr} \hline
Line            &
\multicolumn{1}{c}{$F$($\lambda$)$^{\rm a}$}  &
\multicolumn{1}{c}{100$\times$$F$($\lambda$)/$F$(H$\beta$)}&
\multicolumn{1}{c}{$I$($\lambda$)$^{\rm a}$}  &
\multicolumn{1}{c}{100$\times$$I$($\lambda$)/$I$(H$\beta$)}&
$EW$($\lambda$)$^{\rm b}$ \\ \hline
3727 [O {\sc ii}]        & 228.1$\pm$\, 7.1& 21.87$\pm$0.68& 468.0$\pm$14.9  & 25.82$\pm$0.82& 23.95$\pm$0.45 \\ 
3798 H10                 &  49.1$\pm$\, 4.4&  4.71$\pm$0.42& 114.4$\pm$12.1  &  6.31$\pm$0.67&  6.12$\pm$0.40 \\ 
3835 H9                  &  60.4$\pm$\, 6.3&  5.79$\pm$0.60& 138.1$\pm$16.3  &  7.62$\pm$0.90&  6.67$\pm$0.62 \\ 
3868 [Ne {\sc iii}]      & 218.9$\pm$10.2  & 20.99$\pm$0.98& 438.4$\pm$20.7  & 24.19$\pm$1.14& 24.21$\pm$1.10 \\ 
3889 He {\sc i}+H8       & 175.8$\pm$13.4  & 16.86$\pm$1.28& 365.8$\pm$28.8  & 20.18$\pm$1.59& 21.73$\pm$1.20 \\ 
3968 [Ne {\sc iii}]+H7   & 225.0$\pm$\, 3.8& 21.57$\pm$0.36& 458.4$\pm$10.3  & 25.29$\pm$0.57& 26.58$\pm$0.20 \\ 
4026 He {\sc i}          &  15.3$\pm$\, 1.3&  1.47$\pm$0.12&  29.9$\pm$\, 2.4&  1.65$\pm$0.13&  1.73$\pm$0.13 \\ 
4101 H$\delta$           & 268.3$\pm$\, 4.4& 25.72$\pm$0.42& 532.9$\pm$11.4  & 29.40$\pm$0.63& 30.79$\pm$0.24 \\ 
4340 H$\gamma$           & 462.8$\pm$\, 7.7& 44.37$\pm$0.74& 870.9$\pm$15.4  & 48.05$\pm$0.85& 79.26$\pm$0.57 \\ 
4363 [O {\sc iii}]       & 102.4$\pm$\, 3.7&  9.82$\pm$0.35& 190.0$\pm$\, 6.9& 10.48$\pm$0.38& 19.32$\pm$0.51 \\ 
4471 He {\sc i}          &  35.8$\pm$\, 1.5&  3.43$\pm$0.14&  65.3$\pm$\, 2.7&  3.60$\pm$0.15&  6.57$\pm$0.24 \\ 
4686 He {\sc ii}         &  36.9$\pm$\, 1.8&  3.54$\pm$0.17&  65.4$\pm$\, 3.3&  3.61$\pm$0.18&  6.70$\pm$0.26 \\ 
4740 [Ar {\sc iv}]       &   9.3$\pm$\, 1.1&  0.89$\pm$0.11&  16.3$\pm$\, 2.0&  0.90$\pm$0.11&  1.76$\pm$0.24 \\ 
4861 H$\beta$            &1043.0$\pm$15.3  &100.00$\pm$1.47&1812.5$\pm$27.0  &100.00$\pm$1.49&213.40$\pm$0.61 \\ 
4921 He {\sc i}          &  12.0$\pm$\, 1.6&  1.15$\pm$0.15&  20.5$\pm$\, 2.7&  1.13$\pm$0.15&  2.58$\pm$0.41 \\ 
4959 [O {\sc iii}]       &1078.0$\pm$15.6  &103.36$\pm$1.50&1842.0$\pm$26.8  &101.63$\pm$1.48&204.80$\pm$0.27 \\ 
5007 [O {\sc iii}]       &3214.0$\pm$46.4  &308.15$\pm$4.45&5458.5$\pm$79.4  &301.16$\pm$4.38&576.20$\pm$0.34 \\ 
5876 He {\sc i}          & 115.9$\pm$\, 1.8& 11.11$\pm$0.17& 178.2$\pm$\, 2.9&  9.83$\pm$0.16& 27.80$\pm$0.16 \\ 
6300 [O {\sc i}]         &   8.6$\pm$\, 0.6&  0.82$\pm$0.06&  12.5$\pm$\, 0.9&  0.69$\pm$0.05&  2.36$\pm$0.15 \\ 
6312 [S {\sc iii}]       &   6.6$\pm$\, 0.5&  0.63$\pm$0.05&   9.6$\pm$\, 0.9&  0.53$\pm$0.05&  1.69$\pm$0.12 \\ 
6563 H$\alpha$           &3447.9$\pm$49.8  &330.58$\pm$4.77&4972.8$\pm$78.5  &274.36$\pm$4.33&689.40$\pm$0.30 \\ 
6583 [N {\sc ii}]        &  12.4$\pm$\, 0.5&  1.19$\pm$0.05&  17.9$\pm$\, 0.7&  0.99$\pm$0.04&  2.48$\pm$0.09 \\ 
6678 He {\sc i}          &  31.8$\pm$\, 0.7&  3.05$\pm$0.07&  45.5$\pm$\, 1.1&  2.51$\pm$0.06&  7.46$\pm$0.15 \\ 
6717 [S {\sc ii}]        &  26.8$\pm$\, 0.6&  2.57$\pm$0.06&  38.1$\pm$\, 0.9&  2.10$\pm$0.05&  6.38$\pm$0.14 \\ 
6731 [S {\sc ii}]        &  21.8$\pm$\, 0.7&  2.09$\pm$0.07&  30.8$\pm$\, 1.1&  1.70$\pm$0.06&  5.22$\pm$0.16 \\ 
7065 He {\sc i}          &  49.1$\pm$\, 1.0&  4.71$\pm$0.10&  68.0$\pm$\, 1.6&  3.75$\pm$0.09& 18.59$\pm$0.28 \\ 
7136 [Ar {\sc iii}]      &  21.1$\pm$\, 0.8&  2.02$\pm$0.08&  29.0$\pm$\, 1.3&  1.60$\pm$0.07&  7.62$\pm$0.29 \\ 
7281 He {\sc i}          &   7.6$\pm$\, 0.7&  0.73$\pm$0.07&  10.3$\pm$\, 0.9&  0.57$\pm$0.05&  2.81$\pm$0.29 \\ \\
$C$(H$\beta$)           &\multicolumn{5}{c}{0.240$\pm$0.019} \\
EW$_{abs}$$^{\rm b}$     &\multicolumn{5}{c}{0.9$\pm$0.4} \\
\hline
\end{tabular}

$^{\rm a}$In units 10$^{-16}$ erg s$^{-1}$ cm$^{-2}$.

$^{\rm b}$In \AA.

\end{table*}



\end{document}